\documentclass[twocolumn]{aastex701}
\newcommand\Chandra{\textit{Chandra}}
\newcommand\Copernicus{\textit{Copernicus}}
\newcommand\EUVE{\textit{EUVE}}
\newcommand\FUSE{\textit{FUSE}}
\newcommand\GALEX{\textit{GALEX}}
\newcommand\HST{\textit{HST}}
\newcommand\IUE{\textit{IUE}}
\newcommand\JWST{\textit{JWST}}
\newcommand\Kepler{\textit{Kepler}}
\newcommand\Spitzer{\textit{Spitzer}}
\newcommand\TESS{\textit{TESS}}
\newcommand\XMM{\textit{XMM}}

\received{Nov 25, 2025}
\revised{Dec 17, 2025}
\submitjournal{PASP}

\begin{document}

\title{The Impact of the MAST Data Archive}

\author[0000-0003-4058-5202]{Richard A. Shaw}
\affiliation{Space Telescope Science Institute, 3700 San Martin Drive, Baltimore, MD 21218, USA}
\email[show]{shaw@stsci.edu}
\correspondingauthor{Richard A. Shaw}

\author[0000-0002-8523-015X]{Jenny L. Novacescu1}
\affiliation{Space Telescope Science Institute, 3700 San Martin Drive, Baltimore, MD 21218, USA}
\email{jnovacescu@gmail.com}

\author[0000-0002-4575-9913]{Sarah Weissman}
\affiliation{Space Telescope Science Institute, 3700 San Martin Drive, Baltimore, MD 21218, USA}
\email{sweissman@stsci.edu}

\author[0000-0002-2580-3614]{Travis A. Berger}
\affiliation{Space Telescope Science Institute, 3700 San Martin Drive, Baltimore, MD 21218, USA}
\email{tberger@stsci.edu}

\author[0000-0002-9314-960X]{Clara E. Brasseur}
\affiliation{Lowell Observatory, 1400 West Mars Hill Rd, Flagstaff, AZ 86001, USA}
\email{cbrasseur@lowell.edu}

\author{Jeff Chamblee}
\affiliation{Space Telescope Science Institute, 3700 San Martin Drive, Baltimore, MD 21218, USA}
\email{jchamblee@stsci.edu}

\author[0000-0002-4289-7923]{Brian Cherinka}
\affiliation{Space Telescope Science Institute, 3700 San Martin Drive, Baltimore, MD 21218, USA}
\email{bcherinka@stsci.edu}

\author[0000-0002-9879-3904]{Zachary R. Claytor}
\affiliation{Space Telescope Science Institute, 3700 San Martin Drive, Baltimore, MD 21218, USA}
\email{zclaytor@stsci.edu}

\author[0000-0002-8946-6389]{Theresa Dower}
\affiliation{Space Telescope Science Institute, 3700 San Martin Drive, Baltimore, MD 21218, USA}
\email{dower@stsci.edu}

\author{Chinwe Edeani}
\affiliation{Space Telescope Science Institute, 3700 San Martin Drive, Baltimore, MD 21218, USA}
\email{cedeani@gmail.com}

\author[0000-0003-0556-027X]{Scott W. Fleming}
\affiliation{Space Telescope Science Institute, 3700 San Martin Drive, Baltimore, MD 21218, USA}
\email{fleming@stsci.edu}

\author[0000-0002-8722-9806]{Jonathan R. Hargis}
\affiliation{Space Telescope Science Institute, 3700 San Martin Drive, Baltimore, MD 21218, USA}
\email{jhargis@stsci.edu}

\author[0000-0003-2025-3585]{Julie Imig}
\affiliation{Space Telescope Science Institute, 3700 San Martin Drive, Baltimore, MD 21218, USA}
\email{jimig@stsci.edu}

\author{Tim Januario}
\affiliation{Space Telescope Science Institute, 3700 San Martin Drive, Baltimore, MD 21218, USA}
\email{tjanuario@stsci.edu}

\author[0000-0002-3849-4777]{Karen Levay}
\affiliation{Space Telescope Science Institute, 3700 San Martin Drive, Baltimore, MD 21218, USA}
\email{levay.karen@gmail.com}

\author{Tim Kimball}
\affiliation{Space Telescope Science Institute, 3700 San Martin Drive, Baltimore, MD 21218, USA}
\email{kimball@stsci.edu}

\author[0000-0002-1294-7862]{Jenn Kotler}
\affiliation{Space Telescope Science Institute, 3700 San Martin Drive, Baltimore, MD 21218, USA}
\email{jkotler@stsci.edu}

\author[0000-0002-7871-085X]{Hannah M. Lewis}
\affiliation{Space Telescope Science Institute, 3700 San Martin Drive, Baltimore, MD 21218, USA}
\email{hanlewis528@gmail.com}

\author[0000-0002-4636-7348]{Steve Lubow}
\affiliation{Space Telescope Science Institute, 3700 San Martin Drive, Baltimore, MD 21218, USA}
\email{lubow@stsci.edu}

\author[0000-0003-4827-9402]{Adrian Lucy}
\affiliation{Space Telescope Science Institute, 3700 San Martin Drive, Baltimore, MD 21218, USA}
\email{alucy@stsci.edu}

\author[0000-0002-8058-643X]{Brian McLean}
\affiliation{Space Telescope Science Institute, 3700 San Martin Drive, Baltimore, MD 21218, USA}
\email{mclean@stsci.edu}

\author[0009-0006-9310-9209]{Sunita G. Malla}
\affiliation{Space Telescope Science Institute, 3700 San Martin Drive, Baltimore, MD 21218, USA}
\email{smalla@stsci.edu}

\author{Jacob Matuskey}
\affiliation{Space Telescope Science Institute, 3700 San Martin Drive, Baltimore, MD 21218, USA}
\email{jmatuskey@stsci.edu}

\author[0000-0002-6909-9753]{Sophie J. Miller}
\affiliation{Space Telescope Science Institute, 3700 San Martin Drive, Baltimore, MD 21218, USA}
\email{smiller@stsci.edu}

\author[0000-0001-7106-4683]{Susan E. Mullally}
\affiliation{Space Telescope Science Institute, 3700 San Martin Drive, Baltimore, MD 21218, USA}
\email{smullally@stsci.edu}

\author[0000-0002-7743-8129]{Claire E. Murray}
\affiliation{Space Telescope Science Institute, 3700 San Martin Drive, Baltimore, MD 21218, USA}
\email{cmurray1@stsci.edu}

\author[0000-0003-4797-7030]{J. E. G. Peek}
\affiliation{Space Telescope Science Institute, 3700 San Martin Drive, Baltimore, MD 21218, USA}
\email{jegpeek@stsci.edu}

\author{Carlita Phillip}
\affiliation{Space Telescope Science Institute, 3700 San Martin Drive, Baltimore, MD 21218, USA}
\email{carlita.phillip@gmail.com}

\author[0000-0002-9946-4731]{Marc Rafelski}
\affiliation{Space Telescope Science Institute, 3700 San Martin Drive, Baltimore, MD 21218, USA}
\email{mrafelski@stsci.edu}

\author[0000-0003-1286-5231]{David R. Rodriguez}
\affiliation{Space Telescope Science Institute, 3700 San Martin Drive, Baltimore, MD 21218, USA}
\email{drodriguez@stsci.edu}

\author[0000-0002-4226-304X]{Gregory F. Snyder}
\affiliation{Space Telescope Science Institute, 3700 San Martin Drive, Baltimore, MD 21218, USA}
\email{snyder.gregoryf@gmail.com}

\author[0009-0009-1728-0372]{Achu J. Usha}
\affiliation{Space Telescope Science Institute, 3700 San Martin Drive, Baltimore, MD 21218, USA}
\email{ajusha@stsci.edu}

\author[0000-0002-9194-2807]{Richard L. White}
\affiliation{Space Telescope Science Institute, 3700 San Martin Drive, Baltimore, MD 21218, USA}
\email{rlw@stsci.edu}

\author[0000-0002-4168-239X]{Jinmi Yoon}
\affiliation{Space Telescope Science Institute, 3700 San Martin Drive, Baltimore, MD 21218, USA}
\email{jyoon@stsci.edu}


\begin{abstract}

The Barbara A. Mikulski Archive for Space Telescopes (MAST) hosts science-ready data products from over twenty NASA missions, plus community-contributed data collections, and other select surveys. The data support forefront research in the ultraviolet, optical, and near-infrared wavelength bands. We have constructed bibliographies for each mission from publications in nearly 40 professional journals, and have identified more than 37,000 refereed articles where investigators made a science usage of data hosted in MAST. The publication rate over the last 50 years shows that most MAST missions have had very high productivity during their in-service lifetimes, and have remained so for years or decades afterward. Annual citations to these publications, a measure of impact on research, are robust for most missions, with citations that grow over more than a decade. Most of the citations come from about 10\% of articles within each mission. 

We examined the bibliographies of the active missions \HST\ and \JWST\ in greater detail. For \HST\ the rate of archival publications exceeded those authored by the original observing teams within a decade of launch, and is now more than 3 times higher. Early indications hint that \JWST\ archival articles could dominate the publication rate even sooner. The production of articles resulting from any given observing program can extend for decades. Programs with small and very large allocations of observing time tend to be particularly productive per unit of observing time. For \HST\ in general, a first publication appears within 1.5 yr for 50\% of observing programs, and within 3.8 yr for 80\% of programs. We discuss various external factors that affect publication metrics, their strengths and limitations for measuring scientific impact, and the challenges of making meaningful comparisons of publication metrics across missions. 

\end{abstract}

\keywords{
\uat{Astronomical reference materials}{90} --- 
\uat{Astronomy databases}{83} --- 
\uat{Catalogs}{205} --- 
\uat{Classification}{1907} --- 
\uat{Observatories}{1147} --- 
\uat{Surveys}{1671}
}

\section{Introduction} \label{sec:Intro}

The Space Telescope Science Institute (STScI) hosts one of the largest digital archives of astronomical data in the world. Originally created to be the archive for the \textit{Hubble Space Telescope} (\HST), it was expanded in 1997 to include legacy data from the \textit{International Ultraviolet Explorer} (\IUE) satellite. At that time it was re-named the Multi-mission Archive at Space Telescope to reflect its designation as NASA's permanent archive for space-based missions in the ultraviolet, optical, and near-infrared wavelength bands. In 2012 April it was again re-named, to the Barbara A. Mikulski Archive for Space Telescopes (MAST) to honor the decades-long advocacy of space science from the former U.S. Senator from Maryland. MAST now hosts data and services from over two dozen NASA and select ground-based observatories (see Appendix~\ref{sec:mastMissions}), plus hundreds of curated data collections of high-level science products (HLSPs) that were contributed by the research community.

The scientific accomplishments \added{that were made using data from} 
\HST, the \textit{James Webb Space Telescope} (\JWST), and other MAST-hosted missions have been described over many years throughout the literature. Those papers and others that review scientific discoveries directly attributable to \HST\ or other missions, are perhaps the most effective way to measure impact. 
The promise of such discoveries was the \textit{prima facie} reason these observatories were flown. Yet complications with such an accounting arise, for example, with attribution when more than one mission or observatory contributes to a discovery, or when the impact of a discovery is not appreciated for a long while. In other cases, important and long-standing questions take many years and many observations to reach a conclusion; there may even be controversy over whether certain discoveries are considered settled. Such issues can make any tally of mission discoveries seem incomplete and anecdotal. Bibliometric studies of the scientific literature, on the other hand, offer complementary but more quantifiable and objective assessments, and more nuanced characterizations of the breadth and depth of mission impact. 
Several authors have compiled bibliographies for various space- and ground-based missions, with highly informative results (see \ref{subsec:classification}). 
A few others have characterized the productivity and impact of the body of literature that the research community has produced from the \HST\ mission \citep[e.g.][]{Meylan04, White09, Apai10}. But these last works were published long ago, when important trends had yet to emerge.

The data collections hosted at MAST have accumulated over more than 50 years, and continue to support forefront research based upon active and legacy mission archives. In aggregate, data hosted at MAST consist of a variety of product types (images, spectra, light curves, catalogs, hyperspectral cubes, etc.), span large areas of sky and decades of time, and currently include over 4 PB of data. We refer to data from a given observatory or a major survey as a \textit{mission} (see Appendix~\ref{sec:mastMissions} for a list of missions considered here). Of special note are data collections that were contributed by community science teams. Those high-level science products (HLSPs) typically combine data from multiple observations, programs, or even observatories to derive advanced products, such as deep image mosaics, photometric catalogs, phased light curves, deep and/or time-resolved spectra, spatial maps of ionization structure or extinction, etc. The requirement for hosting these collections in MAST is that the contributing team must have published a refereed paper describing their construction and demonstrating their scientific use, and that the team grants MAST permission to make the data available to anyone, subject to an open-data license that requires proper attribution. These HLSP collections are widely used for follow-on science, and the papers that describe them tend to be highly cited. 

There are a few key attributes of mission data collections hosted at MAST, which in their sum boost the potential for scientific re-use of the data by other investigators. First, the data products are science-ready, which is to say that instrument signature has been removed, the data are fully calibrated (geometrically, radiometrically, temporally, etc.), and the provenance information and documentation are complete and accurate. For missions where STScI conducts science operations (currently \HST\ and \JWST; we refer to these as \textit{Flagship} missions), raw or other lower-level products are offered in addition. The Flagship missions offer open-source calibration software to enable researchers to tune the data processing to their particular science needs. Second, the data are made available in a timely way to the public for research use without restriction (although attribution to the mission is requested and, in some cases, obligatory). For missions where the science programs are determined by competitive selection of observing proposals there is commonly an exclusive access period (EAP; generally one year or less) granted to the investigating teams before the associated data become available to the public. Survey-based missions typically release all data for public use at the end of a survey, or in increments during the course of the survey. Third, for community-contributed data, the collections and their scientific use must be described in one or more refereed publications. Most data at MAST may be searched and retrieved anonymously, apart from EAP data and certain specialized data services which require user authentication. 

We present here a bibliometric study of the more than 37,000 refereed science articles published over the past 50 years to understand quantitatively the productivity, impact, and cross-mission synergy of the missions hosted at MAST. We have identified the extensive set of research publications for each mission using the resources of the NASA Astrophysics Data System (ADS\footnote{ADS Website: \url{https://ui.adsabs.harvard.edu/}}). 
For missions where STScI manages science operations (currently \HST\ and \JWST) we have in addition combined the results of our literature search with information in internal observing program databases to extend our bibliometric analysis to a higher level of detail. This paper describes in Section~\ref{sec:corpora} the domain of the scientific literature surveyed to assemble our bibliographies, the classification system we employed for characterizing the type and extent to which data from MAST missions were used, and an assessment of the accuracy and completeness of our methodology. In Section~\ref{sec:impact} we explore various bibliographic measures of the productivity and impact of MAST-hosted missions over time, with particular emphasis on \HST. We assess in Section~\ref{sec:discussion} the various metrics by which one can measure the impact of MAST, and compare our results with similar bibliometric studies of other archives. We conclude in Section~\ref{sec:conclusions} with an assessment of the value of mission bibliographies in quantifying the impact of major investments in community observing facilities, the value of science-ready data in enabling research, and the impact of NASA mission data on the discipline of astronomy. 

\section{Mission Corpora} \label{sec:corpora}

The corpus of publications for each mission was built up over many years by the staff of the STScI Library and MAST. We searched an extensive list of refereed journals to identify relevant articles for each mission. The processes for reviewing the literature and storing article metadata have improved over time as the supporting technologies matured. Fundamentally, we first execute an automated search of full-text articles to match a set of keywords or phrases that indicate candidate papers of interest, followed by a review by a staff member some months later. (The lag for detailed human review helps keep this activity reasonably efficient, since it is not a full-time endeavor.) We deliberately review only the final journal article, since significant revisions sometimes occur after articles appear in preprint form on arXiv. The human review is necessary because not all authors acknowledge the source of the data, the mission observing program identifier(s) (if relevant), and the MAST archive. \cite{Scire22} also noted the frequent lack of proper facility and data attribution while accumulating \Spitzer\ publications statistics. Acknowledging facilities and including digital object identifiers (DOIs) for the data analyzed is now considered a best practice for data publications in journals \citep{Chen22, D'Abrusco24}. Suggested text for MAST mission and data acknowledgments is provided on our website\footnote{MAST publication resources: \url{https://archive.stsci.edu/publishing}}.

\subsection{Publication Statistics} \label{sec:pubstats}

\begin{figure}[h]
\plotone{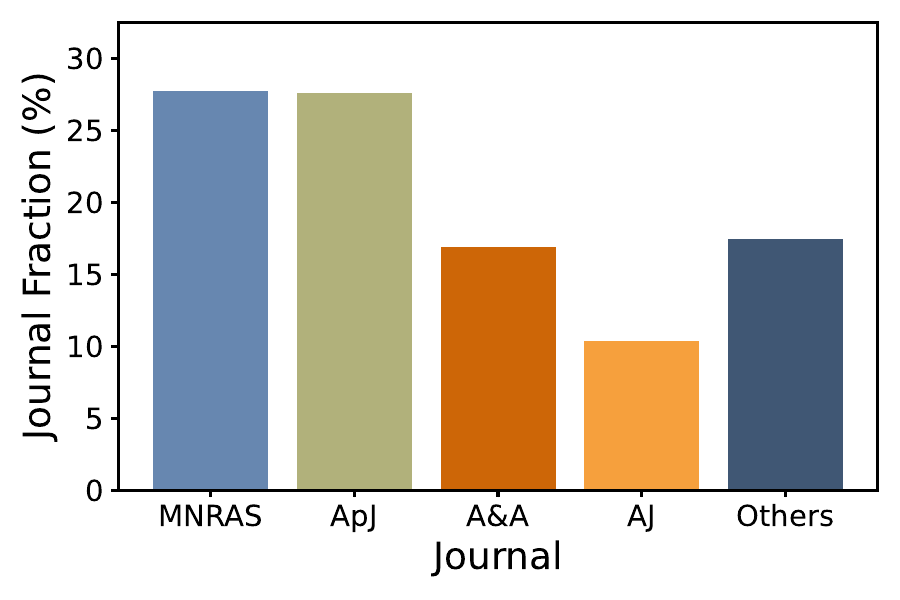}
\caption{Fraction of articles from refereed journals where genuine mission identifiers were encountered during the period 2018~Jan and 2023~May. The categories for ApJ and A\&A both include the main journal, Letters, and the Supplements. The four journals in the axis labels account for more than 80\% of all papers in which a MAST mission name appeared.
\label{fig:papersByJournal}}
\end{figure}

Our current processes were last revised in 2018, and are based upon custom-built software that, among other things, queries ADS to search nearly 40 journals that regularly publish refereed articles on astrophysical research (the list of journals has evolved a little over time). For each published article we record various attributes, including the journal and the publication date, the list of authors, and the number of citations to the article. The searches are performed on a roughly weekly cadence, and cover a rolling interval of time to capture both new articles and metadata updates for previously known articles. The searches match predefined keywords for each hosted mission, and return metadata and links to source journal articles which are stored in our local database, \textit{PaperTrack}. We use multiple keywords per mission to account for the use of synonyms and alternate names for missions (e.g.: \HST, \textit{Hubble}, \textit{Hubble Space Telescope}). The full list of MAST missions and selected abbreviations and alternate names is given in Appendix A. The PaperTrack database includes nearly 59,000 articles published prior to 2023 May, the period through which our literature review is complete for most MAST-hosted missions. It includes an additional 6300 \HST\ and \JWST\ mission articles from 2023 May through 2024 December, where our literature review is more current. 
Figure~\ref{fig:papersByJournal} shows the fraction of articles published after 2018 January from refereed journals. While the fraction from each journal has evolved somewhat over the 50 yr lifetime of MAST missions \citep[see an earlier distribution in][]{Apai10}, we refer hereafter to those that have consistently had most query matches (i.e., MNRAS, ApJ, A\&A, and AJ) as the ``main journals." 

\begin{figure}[h]
\plotone{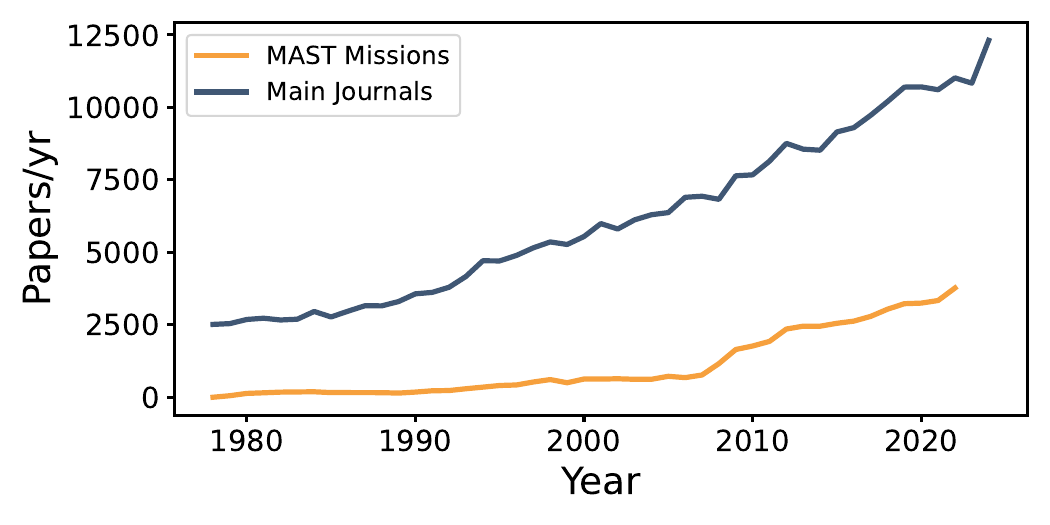}
\caption{Number of papers per year where one or more MAST mission names or keywords appear. Also shown is the growth in the number of refereed papers published per year in the four main journals noted in Figure~\ref{fig:papersByJournal}. 
\label{fig:papersByYear}}
\end{figure}

Currently the text from roughly 8000 articles each year are found to match keywords for some MAST mission, of which about 3500 genuinely refer to a mission. This rate continues to grow as researchers publish new findings from \JWST\ observations, and as they anticipate science that will be enabled with the \textit{Roman Space Telescope}. Interestingly, the number of articles in the main journals has quadrupled over the past five decades, which likely has multiple explanations, including the growth in the number of active researchers in the field. Figure~\ref{fig:papersByYear} shows that papers in the main journals that call out one or more MAST-hosted missions now constitute about one-third of the total published.

\subsection{Classification of Data Usage}\label{subsec:classification}

New articles are reviewed periodically by MAST staff members in order to classify the type of data usage, with the results also recorded in PaperTrack. Since any paper may refer to more than one MAST mission, we refer to the appearance of a mission name or acronym in the text as a \textit{call-out}, and assign a classification for each mission name appearing in a paper by evaluating all call-outs for each mission. The staff review generally consists of conducting a full-text search for mission keywords, stepping through the call-outs until the nature of the data usage is apparent, and then recording the classification for each identified mission in the database. We assign classifications in much the same way as \cite{Apai10} did for \HST, and apply the following criteria:

\textbf{Science.} The authors reached one or more scientific conclusions based upon an original analysis of mission data hosted by MAST. The authors may have used raw or calibrated products, and may or may not have combined or otherwise transformed products prior to analysis. The usage may be direct (e.g., creating a photometric catalog from images) or a quantified non-detection (e.g., ruling out photometric contamination of a source observed with a ground-based telescope based on a non-detection of nearby sources in an \HST\ image). 

\textbf{Mention.} The authors called out a MAST Mission by name or acronym, but did not make direct use of mission products. Often this is in the context of citing prior work that made use of mission data from MAST. Authors who mention a mission in advance of science operations are very often anticipating science that could be accomplished with the mission. Apparent call-outs are ignored entirely if the mission acronym is merely used as a part of a source designation. 

\textbf{Instrument.} The authors described the design, calibration, or performance of a specific instrument on the mission. If the article also evaluated instrument performance based upon a scientific analysis of observations from an astrophysical source, it is instead classified as Science.

\textbf{Engineering.} The authors described the design of the telescope itself, the instrumentation as a whole, or operational software and other support systems. 

\textbf{Data Influenced.} The authors made indirect use of mission data hosted by MAST, but did not themselves analyze mission data products. For example, including an image from \HST, where the authors have added wire-frame annotations of apertures used to indicate the positions of ground-based spectra within some extended source, would be classified as Data Influenced. 

For internal purposes we also track a category for papers that were ignored by reviewers to account for false-positive call-outs in papers (e.g., the word \Kepler\ is associated with a NASA mission, but it is also the name of a crater on the Moon, a supernova remnant, a software application to model supernova explosions, a term used in stellar or planetary dynamics, an institution, and the name of a famous astronomer). Any given call-out may not indicate a scientific use of mission data, but an indication of science use in any call-out will result in a Science classification. For a given article, each mission mentioned is classified; an article may have only one classification type for each mission. 

\begin{figure}[h]
\plotone{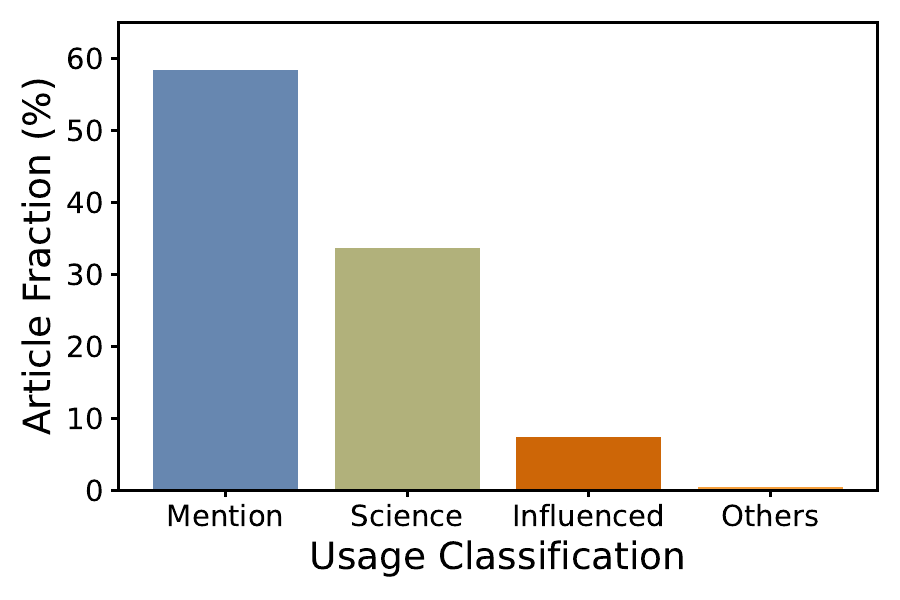}
\caption{Histogram of data usage classification fractions that were assigned after 2018 Jan 01. See text for classification criteria. 
\label{fig:papersByClass}}
\end{figure}

Historically only the first three of the categories were used, and the latter categories were added within the last 10 years. One caveat is that the criteria used for the Science classification evolved somewhat since MAST bibliography reviews first began. That is, in the period prior to 2018 certain papers deemed important in describing the mission, the instrumentation, and reviews of science results were included this category, though today they mightc have been assigned a different classification. The number of such papers is quite small and will not affect the conclusions of this paper. Since 2018 January the categories Mention, Science, and Data Influenced constitute the bulk of the assigned classifications (see Figure~\ref{fig:papersByClass}); those articles that include a Science classification will be the focus of the analysis in this paper. These Science articles for many missions (including \GALEX, \HST, \IUE, \JWST, \Kepler, K2, Pan-STARRS, \TESS) have been incorporated into separate ADS \textit{bibgroups}, allowing users to include a mission corpus in ADS search criteria. The other usage categories undoubtedly contain useful information about impact on the research community (as, by analogy, mentions of commercial products in social media are valuable to on-line companies) but that is a subject for another paper. 

Additional details were harvested for the MAST Flagship missions, for which STScI has responsibility for Science Operations and for producing the science-ready data products that are offered to users through MAST. The information collected for papers classified as Science includes the ID(s) of the observing program(s) that obtained the data, the instrument(s) used, and the specific datasets that were analyzed if obtainable. Other ancillary information is harvested from internal Flagship mission databases, including the time allocation, and whether the paper authors were also investigators of the observing program(s) that obtained the data. For \JWST\ the instrument observing template (i.e., the instrument configuration and the data-taking mode) was also included when possible to discern. This information is for the most part used internally to better understand community demand at a higher level of granularity. 

The methodology we just described for constructing MAST bibliographies --- identifying, reviewing and classifying data usage for a mission --- closely follows the recommendations of \citet{D'Abrusco24}. 
Several authors have compiled bibliographies for other space- and ground-based observatories: e.g., \cite{Crabtree08} for Gemini, \cite{Rots12} for \Chandra; \cite{Savaglio13} for \textit{Swift}; \cite{Scire20, Scire22} for \Spitzer; \cite{Ness25} for \XMM; and \cite{Demarchi24} for various missions supported by the European Space Agency. Most of these bibliographies involve some type of automated search (usually with ADS) of refereed literature, with follow-on human evaluation to select appropriate papers. Most also adopt some variant of the \cite{D'Abrusco24} criteria for science publications, namely that the article must present an original analysis of data from their observatory. Variations from this generally involve including (or not) articles that make use of source catalogs generated from mission data: articles are included in the MAST bibliography where the original catalog is created, but (currently) not publications that make use of that catalog, except if the catalog is closely associated with the Mission \citep[e.g., the Hubble Source Catalog,][]{whitmore16} or is hosted by MAST in an HLSP collection. 

\subsection{Completeness and accuracy}\label{subsec:completeness}

The paper corpora have expanded at an accelerating rate for the past decade. Over 25 staff at STScI have participated at some point in paper reviews during that time, with some division of labor between reviews for Flagship missions vs. all others. Practical experience with the classification scheme described above uncovered a small fraction of cases where the most appropriate classification was ambiguous; often in these cases the paper authors were not clear about how they used data from observing facilities or were not careful with their attributions. Over the past few years some staff have explored the possibility of using machine learning (ML) or artificial intelligence (AI) techniques to classify data usage within research publications \citep[see, e.g.,][and J. Wu, et al. in preparation]{Shaw18, AmadoOlivo25}. Over the course of evaluating ML results it became apparent that some small fraction of the human classifications in our database were not accurate. While identifying flaws in prior classifications is a desired outcome from the ML/AI exploration, we wanted to quantify the nature and extent of the discrepancies for the present work. We formed two independent teams to re-review small, but reasonably representative samples of papers in our PaperTrack database. One paper sample focused on the Flagship missions, and the other focused on the remaining missions. The aims were to measure the random error in human classifications (i.e., from mistakes or inadequate evaluation), measure the systematic error in human classifications (i.e., from different interpretations by different people of the classification criteria), examine the boundaries and ambiguities of the classification criteria, and produce a gold sample of robust classifications for comparison to future efforts to train or evaluate ML/AI techniques. 

The non-Flagship sample consisted of 99 papers with publication dates between 2018 Jan and 2023 May; the papers contained 234 distinct mission call-outs. The Flagship sample consisted of 120 papers with publication dates between 2022 Jan and 2023 Dec to maximize the operational overlap between \HST\ and the just-commissioned \JWST; this sample contained 186 distinct mission call-outs. Both samples included a few papers that were selected specifically because one or more of the classifications had been noted as challenging. But generally the papers were selected to be reasonably representative with respect to the publishing journals, missions, paper classifications, and the fraction of papers that staff members had reviewed in prior years. Between four and six staff members performed evaluations of each mission identified within a paper, depending upon whether it had been evaluated historically by a review team member or by some other staff member. 

\begin{figure}[h]
\plotone{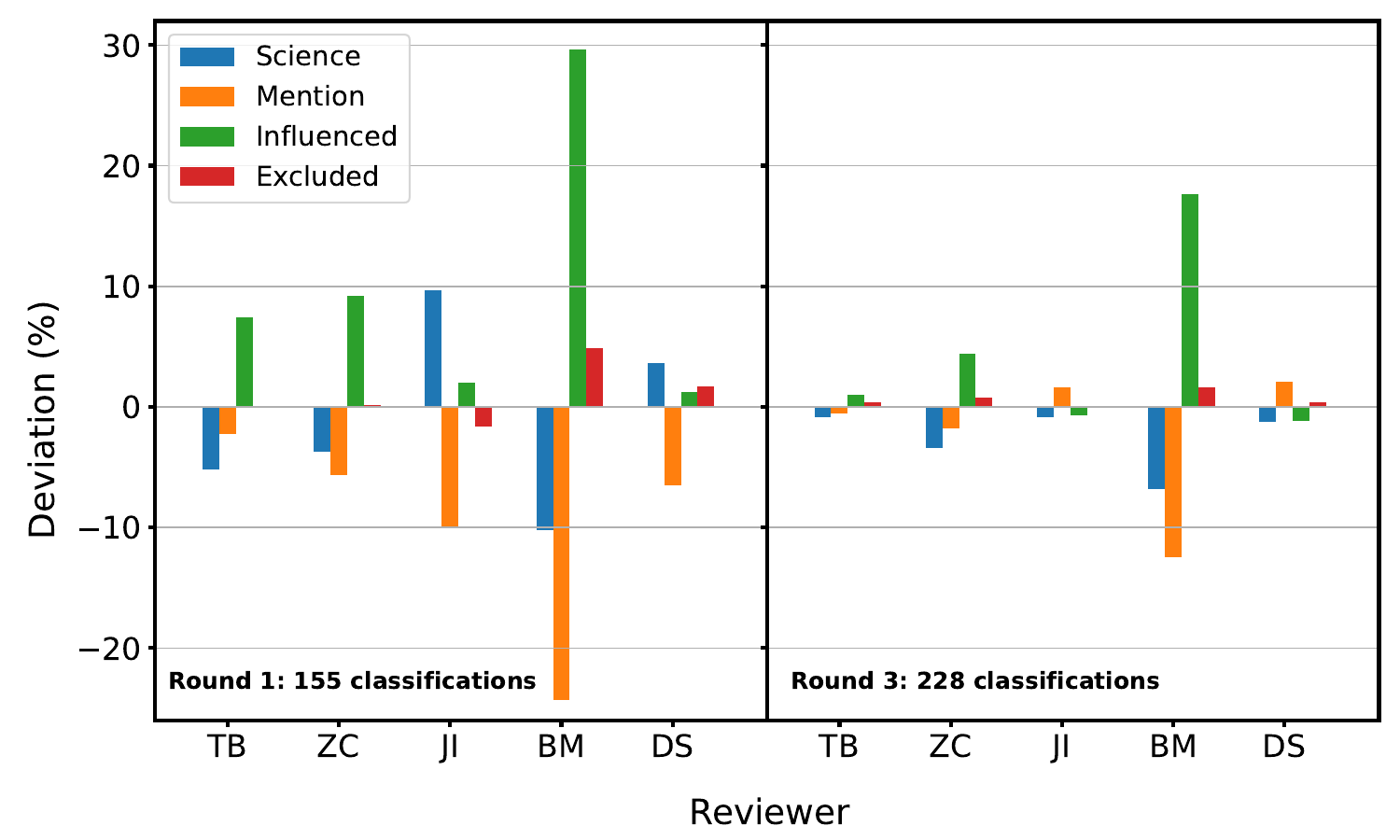}
\caption{Histogram of deviations of paper classifications (colored bars) by team members (designated by initials along the $x-$axis) from the team consensus view of non-Flagship mission papers. The deviations decreased markedly from the first round of re-review (\textit{left}) to the third round (\textit{right}). The deviations should be interpreted something like: "Reviewer X assigned Mention 5.1\% more often than the consensus view, and assigned Science 2.1\% less."
\label{fig:bibReviewDev}}
\end{figure}

The re-review for both samples was conducted in three rounds. In the first round each member reviewed each paper in their team's sample independently, and evaluated the call-outs for each mission following our standard practice. Upon assessing the results for the non-Flagship sample, we found that call-outs for some missions had been missed in the historical evaluations (and indeed, some had been missed by review team members in the first round of the re-evaluation). We also found that the teams had not achieved perfect consensus on the usage classification for some fraction of cases. Consensus was defined for the Flagship review as agreement among 3 of 4 members, and for non-Flagship missions as agreement among 4 of 5 team members. In the second round team members reviewed the mission call-outs that they missed, and re-examined cases where classifications among team members were particularly discrepant. While the rate of consensus in the second round improved, we wondered whether the remaining discrepancies should be ascribed to misunderstandings of the papers, or to honest disagreements about how the classification criteria should apply. Each team met to discuss the remaining discrepancies in a third round of review, for which the aim was not to press for convergence, but to understand the nature of the discrepancies. After round 3, consensus was achieved for 
more than 97\% of the missions over all papers. The discrepancies for the non-Flagship sample after round 1 and round 3 reviews are shown in Figure~\ref{fig:bibReviewDev}. Finally, we updated the PaperTrack database with the consensus classifications where they disagreed with the historical classifications. 

These team re-reviews helped to quantify multiple aspects of the quality of the literature reviews, which are summarized in Table~\ref{tab:completeness}. Based upon the mission call-outs that were missed in the historical paper evaluations, and bearing in mind the small sample size, we estimate the completeness of our database to be approximately 90\% for non-Flagship missions and about 97\% for Flagship missions. Completeness was confirmed with an informal follow-up review of 50 additional papers that had historically been ignored for lacking genuine mission call-outs. For most of the cases where a mission had not been identified in a paper, it would likely not have been classified as a Science, which makes sense since science usages correlate with a larger number of call-outs for a given mission within a paper, making them harder to miss. We have since improved our workflow for paper reviews to address shortcomings in completeness.

\begin{deluxetable}{lcccc}[h] 
\tabletypesize{\footnotesize}
\tablewidth{0pt} 
\tablecaption{Historical Completeness and Classification Accuracy\label{tab:completeness}}
\tablehead{
    \colhead{} & \colhead{} & \colhead{} & \colhead{Science} & \colhead{Science} \\
    \colhead{Sample} &
    \colhead{Papers} &
    \colhead{Mission IDs} &
    \colhead{Completeness} &
    \colhead{Usage} 
}
\startdata
Flagship     & 120 & 186 & 97\% & 94\%\tablenotemark{a} \\
Non-Flagship &  99 & 234 & 90\% & 95\% \\
\enddata
\tablenotetext{}{\JWST\ Science usage was identified $>99$\% of the time.}
\end{deluxetable}

The classification of data usage was more nuanced, however. While the consensus classifications differed from the historical classifications about 7\% (\HST) and 11\% (non-Flagship) of the time, many of the discrepancies were between Mention and Data Influenced. From this we conclude that the definition of Data Influenced needs to be more robust. But the rate of discrepancies between Science and some other category, which is what matters for the analysis in this paper, is approximately 5\%. The historical completeness and classification accuracy for \JWST\ was even more robust: only about 1\% of papers with a \JWST\ call-out were missed, and Science usages were misclassified less than 1\% of the time. Our conclusions are that the corpus for each mission is at least 90\% complete (with better completeness for the Flagship missions), with as much as 95\% classification accuracy for Science usages. Achieving better than 97\% classification accuracy is unlikely to be possible for non-Flagship missions. The productivity and impact metrics described in the following sections will reflect the historical accuracy, while future reviews, or re-assessments of historical reviews aided by ML/AI, will be closer to the achievable accuracy.

\section{MEASURES Of Mission IMPACT\label{sec:impact}}
\subsection{Publications by Mission}

\begin{figure*}[h]
\plotone{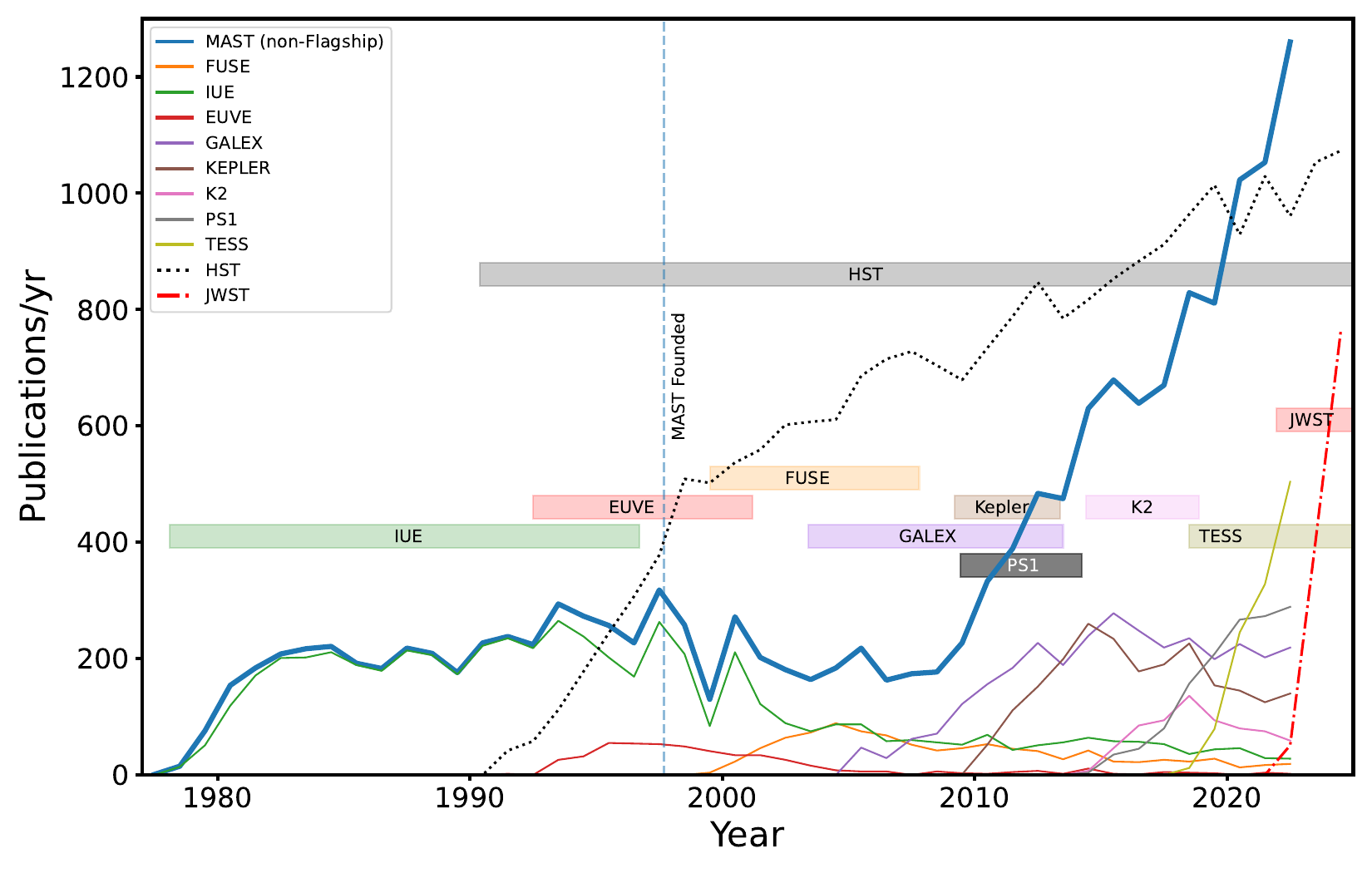}
\caption{Refereed publications per year where data from a MAST-hosted mission was used in the analysis. Curves are color-coded per mission (see legend and Appendix~\ref{sec:mastMissions}); most mission statistics are complete only through 2022. The total of MAST missions called-out in these papers (exclusive of the Flagships: \HST, \JWST) is also shown (thick blue curve). The spans of mission in-service dates are indicated (colored bars). 
\label{fig:mastPubs2025}}
\end{figure*}

The pace of publications in scientific journals is a classic measure of scientific productivity for a mission \citep[e.g.,][]{Apai10, Rots12, Scire22, Demarchi24, Ness25}. Our PaperTrack database records more than 37,000 articles classified as Science that were published prior to 2025 Jan. Figure~\ref{fig:mastPubs2025} shows the number of refereed articles published each year that analyze data from MAST-hosted missions with the largest data volume (the full names for missions included in this study are given in Appendix~\ref{sec:mastMissions}). Note that full-year bibliographies for non-Flagship missions are only complete through 2022. It is clear that mission data are used extensively to advance science, with a solid publication rate for most missions. Interestingly, the pace of \HST\ papers shows a steep rise through 1998 but increases more slowly in the following years, as noted by \cite{Apai10} who attribute the change in slope to the rate of paper production equilibrating to the rate at which new data were being obtained from the mission. The pace of non-Flagship publications increases sharply after 2009, which we attribute primarily to the addition of data from multiple smaller missions to MAST: \GALEX\ and \Kepler\ initially, and later Pan-STARRS and \TESS. Perhaps the most impressive conclusion is that the analysis of mission data continued at a robust rate long after missions have ceased operations. Further, the total publication rate over all non-Flagship missions rivals that of the active Flagship missions. This is a testament to the enduring value of science-ready mission data in astronomy. 

\subsection{Citations to Mission Publications\label{subsec:citesMissionPubs}}

A classic measure of the science impact of a paper is the number of the accumulated citations. We analyzed the incidence of citations over time for MAST mission science papers, although we agree with \cite{Rots12} that it is easy to over-interpret citations in isolation, and comparing citations between missions is fraught. Citations are a trailing indicator of impact, in that it takes some time for a paper to accumulate most of the citations it will ever receive. Yet a broad analysis of citation trends is highly revealing. 

\begin{figure}[h]
\plotone{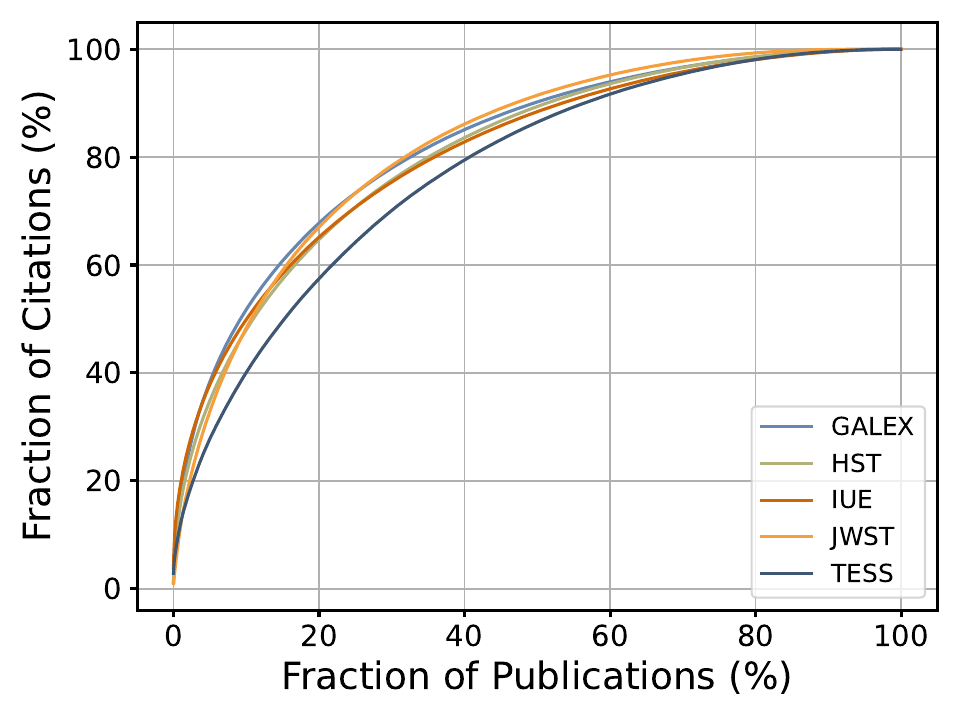}
\caption{Cumulative fraction of citations over refereed science papers that analyzed data from select MAST missions (see legend); the curves for other MAST missions are nearly identical. The distributions seem to follow a 50:10 rule: half of the citations come from the most highly cited 10\% of papers for the mission, while the least cited half of papers account for only 10\% of mission citations. 
\label{fig:citeFract}}
\end{figure}

Prior to 2023, citations to MAST mission science publications totaled 2.0 million, of which 1.4 million were to \HST\ alone (see Appendix~\ref{sec:mastMissions}). Only a few hundred of the 37,000 MAST mission science articles ($\sim1$\%) have no citations at all. Indeed, the median number of citations to individual articles over the corpora described here is quite high: 25 over all missions, and 35 for \HST\ papers alone. Figure~\ref{fig:citeFract} shows the distribution of citations over the articles published prior to 2025, for selected missions. The citations are heavily weighted to a small fraction of articles, almost independently of the mission or of how recently the papers were published. There are a significant number (91 at this writing) of articles with more than 1000 citations each. Two MAST archival research articles have more than 10,000 citations (one uses data from both \IUE\ and \Copernicus), which puts them in an exclusive club (11, at this writing) of the most-cited, refereed articles for all time in the ADS ``astronomy" collection. 

\begin{figure}[h]
\plotone{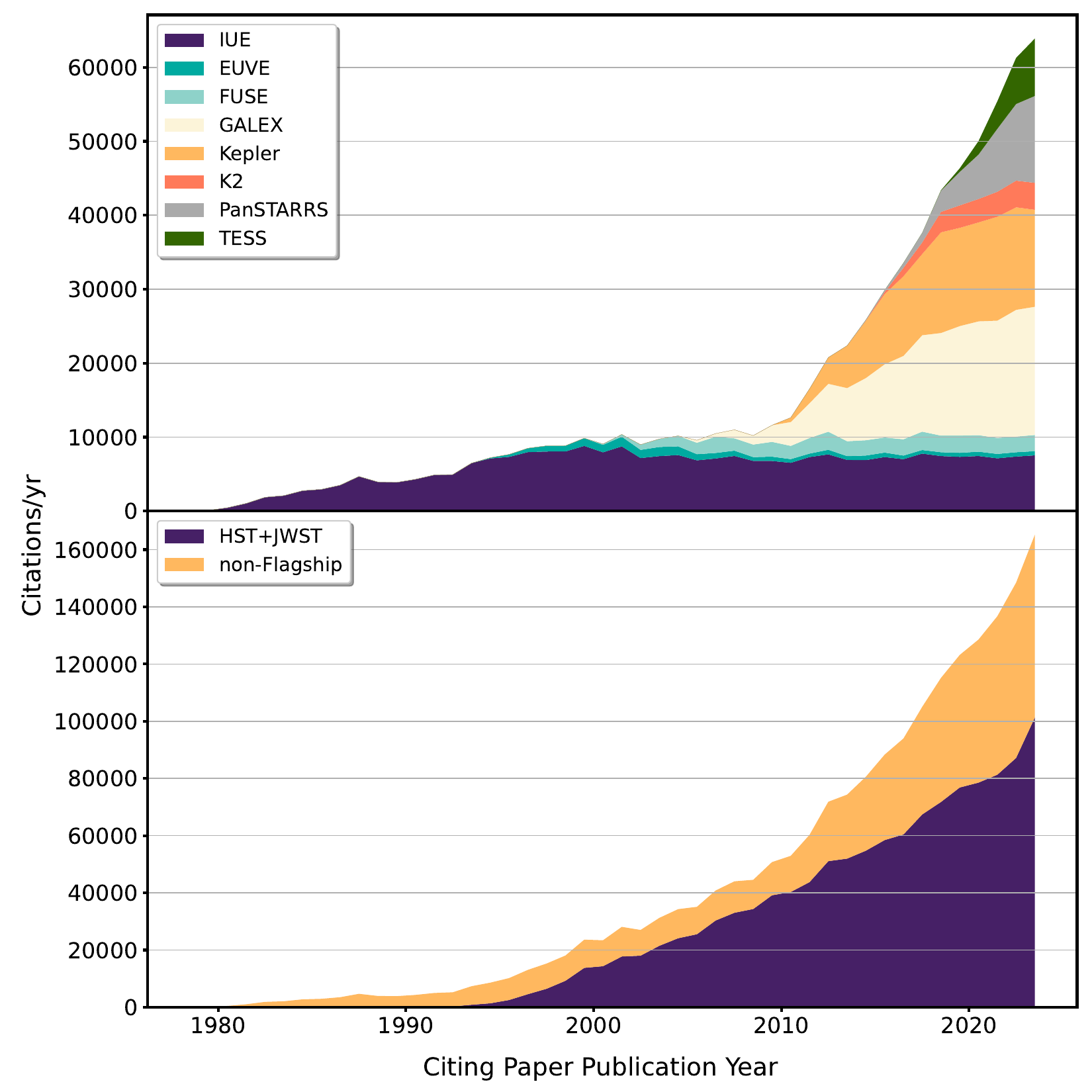}
\caption{Number of citations per year to Flagship and non-Flagship science papers (\textit{lower}), and similarly for a breakdown of most non-Flagship missions (\textit{upper}), with missions indicated (\textit{colored curves}: see legend). Annual citation rates are derived from the publication year of the citing papers.
\label{fig:citesMast}}
\end{figure}

Figure~\ref{fig:citesMast} shows the annual citation rates for the most productive missions. The sharp rise in citations to non-Flagship mission papers after 2009 is the result of adding data from new missions to the MAST archive: initially from \GALEX\ and \Kepler, and later Pan-STARRS and \TESS. Interestingly, the citation rate for a given mission remains substantial after the mission ceases operations, even for long-retired missions, indicating that the results presented in those mission papers are still relevant for current research. 

\subsubsection{Impact of Individual Mission Publications}

\begin{figure}[h]
\plotone{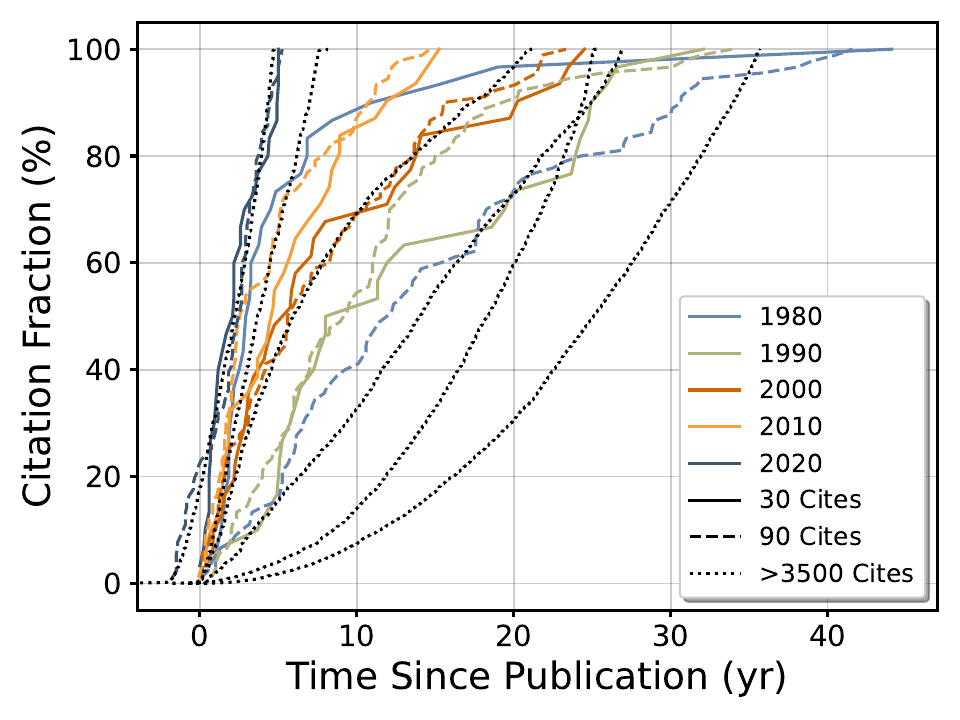}
\caption{Fraction of accumulated citations vs. the elapsed time since publication, or Curves of Growth (CoG), for a sample of article pairs. These example pairs have publication dates from 1980 to 2020 in integral decades (colored curves). The pairs were constructed from articles with 30 citations (solid curves) and with 90 citations (dashed curves), and are displayed in the same color. Also plotted are CoG from any year that have accumulated more than 3500 citations (dotted black curves). Note in particular the curves for the pair published in 1980 (light blue), where the time to reach 80\% of the accumulated citations is separated by decades. The citation growth for the very most highly cited articles are still accelerating, with no sign of a turn-over. 
\label{fig:citesCog}}
\end{figure}

The extent to which individual science articles have lasting impact can be measured by assessing the rate at which papers accumulate citations. We selected example pairs of science articles from MAST missions that had accumulated 30 (a number typical of mission articles) and 90 citations. These pairs had been published in successive decades from 1980 through 2020. We queried ADS to fetch the years of the citing papers, and plotted the fraction of accumulated citations each year since publication (see Figure~\ref{fig:citesCog}). These citation curves of growth (CoG) often show a sharp rise to an inflection within 10--15 years of publication, followed by a turn-over to a lower and steadily decreasing accumulation rate later on. What is perhaps surprising is that the characteristic time interval over which a mission article accumulates 80\% of all accumulated citations varies widely (note the curves in Figure~\ref{fig:citesCog} for the two papers published in 1980), with a period of impact sometimes exceeding 20 years. 
We suspect that these CoG features are not limited to publications related to MAST missions, but apply more broadly. 
The shape of the curves suggests that citations to the articles published in 2010 or later are still accumulating rapidly (i.e., there is little evidence of a turn-over). Also plotted in Figure~\ref{fig:citesCog} are the CoG for selected articles that have accumulated more than 3500 citations. The rates for most of these ``superstar" papers suggest that the citation turn-over is nowhere in sight. The upshot is that impact of MAST mission science papers (and perhaps all refereed science papers in astronomy) is clearly best measured in decades, not years, but predicting which ones will accumulate the largest numbers of citations is difficult based solely on the accumulation rate within $\sim10$ yr of publication. 

It is perhaps also worth considering who is citing MAST mission papers. We found, by exploring the metrics associated with the ADS bibgroups associated with MAST missions (\HST, \JWST, \IUE, etc.) that the top-ranked mission papers by citation rate include some of the most prolific and highly cited researchers in the discipline of astronomy. Many of these researchers have a bibliography that has attracted tens of thousands of citations — more than the citation total for some NASA missions. 

\subsection{Archival Research}

We define archival research, and the resulting publications, as analysis of mission data by investigators other than the General Observer (GO) team who obtained the original observations. In this sense all papers published from survey-focused missions, and some papers published during and after PI-focused mission operations have ceased, might be considered archival papers. From the earliest days of \HST, archival research during the operations phase was expected, and in some cases explicitly funded. Yet it is remarkable that archival research papers came to dominate the \HST\ publication rate so soon. In what follows we leverage our classification of publication types for the Flagship missions to offer additional insight into how archived data were used. 

\begin{figure}[h]
\plotone{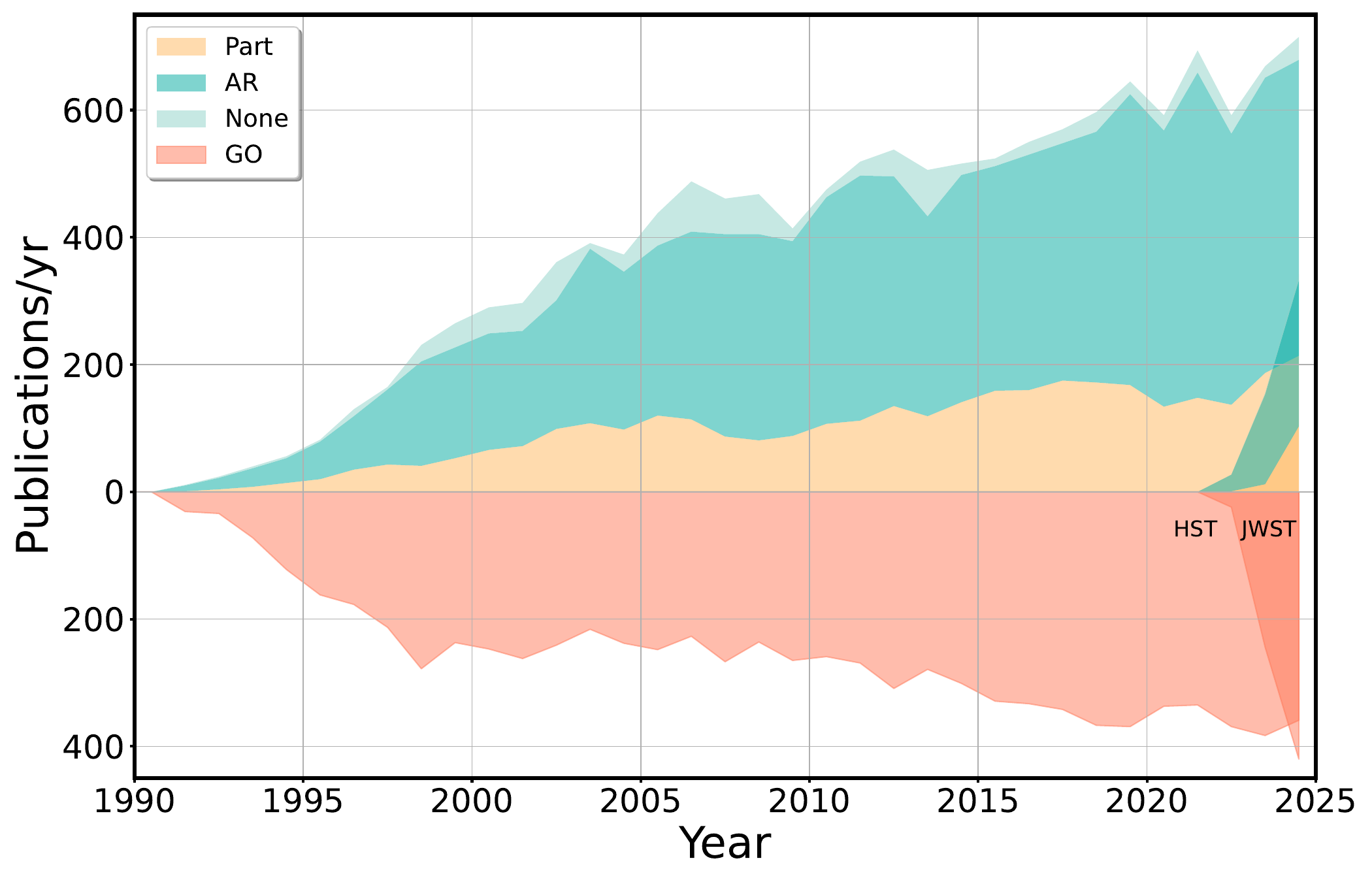}
\caption{Number of papers published per year that analyzed \HST\ (\textit{lighter shades}) and \JWST\ (\textit{darker shades}, at right) data by author category (see legend; ``None" means the author category could not be determined). This split-from-zero plot shows the number of papers authored by the original investigating team (GO) on the lower (below zero) $y-$axis. 
\label{fig:flagPubs}}
\end{figure}

\cite{White09} and \cite{Apai10} noted that the number of \HST\ archival research papers had by 1998 already begun to exceed those produced by the original GO teams. We adopt their definition of an archival paper: that there is no overlap between the paper co-authors and the co-investigators on the observing program that obtained the data. Papers where data from more than one Flagship program were analyzed are part archival if there is no overlap between the paper authors and the investigators of at least one of the observing programs, but there is overlap on any other observing program. This approach to distinguishing archival research, while simple to implement, may well under-count archival papers, in that some GO team members certainly do re-use data they obtained for follow-on research with other collaborators. Figure~\ref{fig:flagPubs} reprises and updates results from \cite{White09} and \cite{Apai10}, showing that papers authored by the original investigating team (``GO" in the figure) were more common in the early years of \HST, but the fraction of GO papers is currently less than half that published by the sum of archival or part-archival researchers (AR). \cite{Apai10} noted the relatively flat production of GO papers from about 1999 through 2010, and suggested that the publication rate was equilibrating to the production of new data from the telescope. Now, 15 years later, we note a steady rise in the production of papers classified as GO after 2010. Several programs with large time allocations (see Sect.~\ref{sec:discussion} and Appendix~\ref{sec:progCategories} for a definition of Large programs) were awarded beginning in 2010, and many of these involve very large teams of co-investigators and result in very large numbers of papers (see Sect.~\ref{subsubsect:prodByProg}). 
We suspect that the increase in GO classifications after 2010 is due at least in part to GO papers that were published a decade or more after the data were obtained, which arguably ought instead to have been counted as archival. Finally, the rapid rise of papers that analyze \JWST\ data is steeper than for any other mission hosted by MAST, and the fraction of non-GO papers (about one-third) is quite large considering that science operations began only in 2022 July, and the default duration of the EAP is one year. 

\subsection{Cross-Mission Publications}

The extent to which paper authors analyze data from multiple missions offers some insight into the cross-mission synergy of archival research. There are a total of 3365 such papers, or about 10\% of the total number of Science papers from MAST missions. Table~\ref{tab:multiMission} lists in the first column the number of missions, $N$, where data were analyzed in the same paper, followed by the corresponding number of papers that analyzed data for $N$ missions, followed by example combinations of missions and their relative fraction of the total from the second column. The number of multi-mission papers and the number of missions analyzed in each paper would both be elevated if the use of data from non-MAST missions were available and included. Most of the possible mission combinations are represented, and most include \HST\ (because most MAST papers analyze data from \HST).

\begin{deluxetable}{crl}[h] 
\tabletypesize{\footnotesize}
\tablewidth{400pt} 
\tablecaption{Papers Analyzing Data from Multiple MAST Missions\label{tab:multiMission}}
\tablehead{
    \colhead{N} &
    \colhead{Papers} &
    \colhead{Common Combinations}
}
\startdata
2 & 2674 & \HST+\IUE: 22\%; \\
  &      & \HST+\JWST: 19\%; \\
  &      & \HST+\GALEX: 19\% \\
3 &  562 & \FUSE+\HST+\IUE: 25\%; \\
  &      & \GALEX+\HST+one of \\
  &      & (\JWST, \IUE, or PanSTARRS): 22\% \\
4 &  109 & \FUSE+\HST+HUT+\IUE: 10\% \\
5 &   15 & \Copernicus+\FUSE+\HST+HUT+\IUE: 20\% \\
6 &    4 & \FUSE+\HST+4 others: 100\% \\
7 &    1 & \FUSE+\GALEX+\HST+HUT+\IUE+ \\
  &      & PanSTARRS+WUPPE \\
\enddata
\end{deluxetable}

\subsection{Publications from Observing Programs}

\begin{figure*}[h]
\plotone{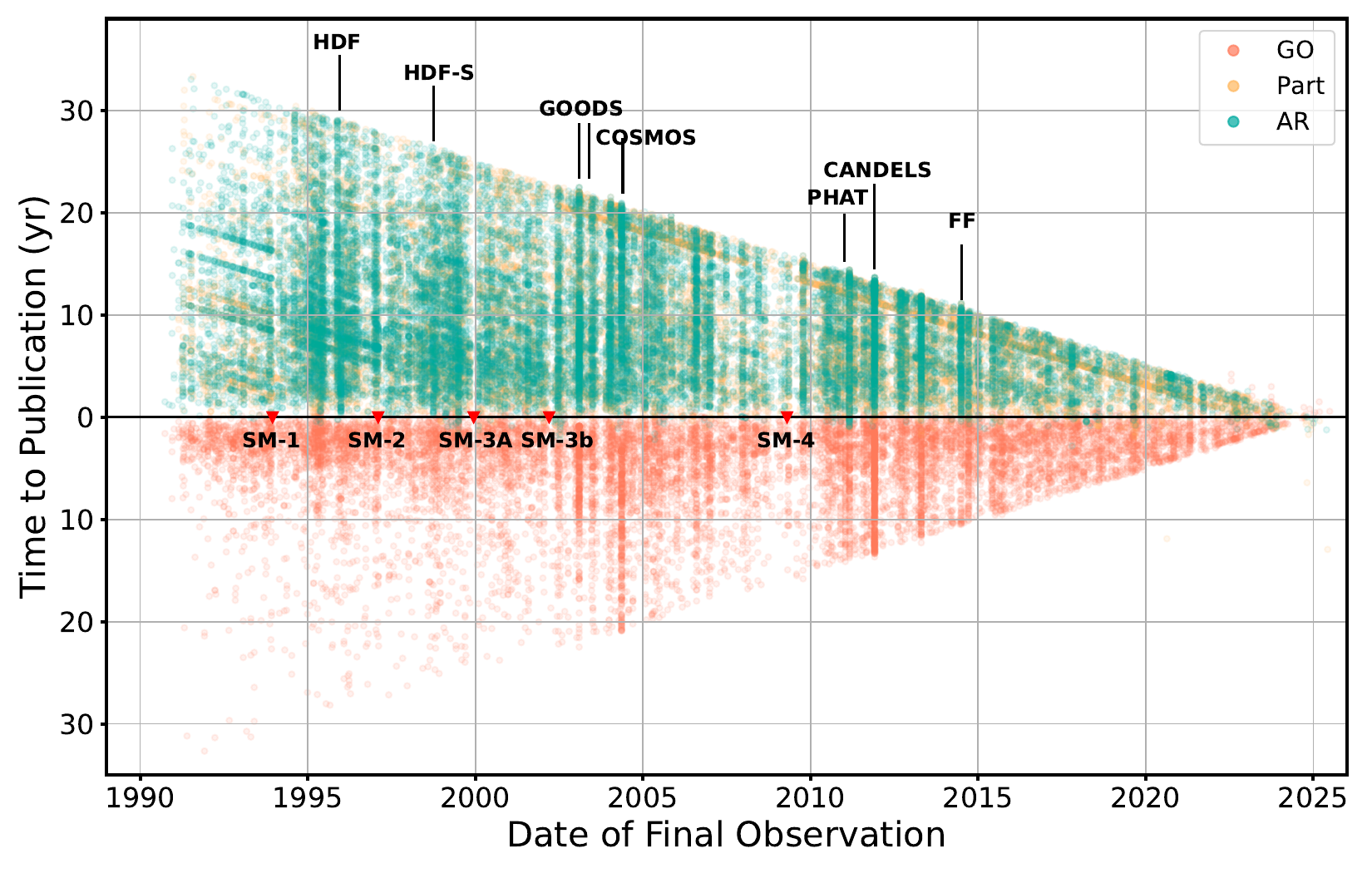}
\caption{The time elapsed from the last observation in an \HST\ observing Program to the date of all refereed publication vs. date of the observation, coded by author category. The lower (below zero) $y-$axis shows category "GO", meaning the proposal team published the article; the upper $y-$axis shows the Archival and Part-archival papers. The dates of the five \HST\ servicing missions are marked (\textit{inverted red triangles}). The names of representative survey programs with large time allocations are also marked: 
the Hubble Deep Field \citep[HDF:][]{Williams96} with 
the follow-on program in the south \citep[HSF-S:][]{Williams00};
the Great Observatories Origins Deep Survey \citep[GOODS:][]{Giavalisco04};
the Cosmic Evolution Survey \citep[COSMOS:][]{Koekemoer07}; 
the Panchromatic Hubble Andromeda Treasury \citep[PHAT:][]{Dalcanton12}; 
the Cosmic Assembly Near-IR Deep Extragalactic Legacy Survey survey \citep[CANDELS:][]{Koekemoer11}; 
and the Frontier Fields survey \citep[FF:][]{Lotz17}.  
\label{fig:obsPubs}}
\end{figure*}

Many observing programs result in multiple, or even hundreds, of publications. We focus for the moment on \HST\ publications, where the substantial quantity of bibliographic data and information about the originating observing programs enables detailed analysis. Figure~\ref{fig:obsPubs} shows more than 50,000 instances where \HST\ data obtained from observing programs were analyzed within refereed papers, with date of the last observation for the program plotted against the time elapsed to the publication, color-coded by the author category from Figure~\ref{fig:flagPubs}. 
\citep[][published a similar plot to show the enduring value of \HST\ datasets.]{Peek17}
Trends in the density of points tell an interesting story. As might be expected, the bulk of the publications by the original observing team (GO) occur within a few years of the completion of observations for their programs. Interestingly, there are some cases where observing teams publish several years, or even decades after the last observation for a program. These later publications may be part of a long-running series of articles that explore multiple aspects of the science that can be gleaned from the data. Alternatively, these may be cases where the article was misclassified as GO because one or more of the co-authors was also a member of the proposing team, rather than a participant in a genuine archival paper (see Sect. 3.3). Publications by archive researchers (AR), in contrast, begin soon after data become available to them (often more than a year after an observation), and often continue for many years afterward. This is particularly true for Large, Treasury, or Director's Discretionary programs (see Appendix~\ref{sec:progCategories} for the definition of program categories) that involve large allocations of time, and for which the data become public as soon as they are archived. Finally certain diagonal features in the figure (with a slope of $-1$) correspond to individual (part-)archival papers that analyzed data from a large number of programs \citep[e.g.][]{Kruk23}.

Pauses in \HST\ science observations are discernible at the epoch of each servicing mission (SM) to \HST. The dearth of observations from late 2008 though mid-late 2009 is especially noteworthy, and corresponds to a time when difficult hardware problems were encountered in the lead-up to SM~4, which was postponed from 2008 Oct to 2009 May to prepare to replace a failed data-handling computer. Difficulties with re-enabling the science instruments during that time, and failure of the NICMOS instrument, substantially disrupted science observations \citep[see][]{Reid09}. The vertical bands of color are from large observing programs, often Director's Discretionary or Treasury programs, for which data were immediately made public. These programs were designed to enable multiple types of follow-on science, and they account for a sizable portion of all the publications (see Sect. 4.3). 

\subsubsection{Time to First Publication}

Investigating teams who were awarded observing time with the Flagship missions enjoy multiple means of support to analyze and publish their science results, including high-speed access to science-ready data products via MAST, highly accurate calibrations, software for optimizing the data processing to their particular science needs and, for US investigators, funding support. Apart from the usual academic incentives to publish their science results promptly, most investigating teams were granted a finite exclusive access period (EAP) for the data they obtained, after which the data become available for anyone to analyze and publish. Note, though, that certain types of programs by policy had a zero-length EAP, and some proposers chose to shorten or waive their EAP to improve the chance that their programs were awarded observing time. Given the support and incentives, we were interested in assessing how rapidly the science results were published. 

\begin{figure*}[h]
\plotone{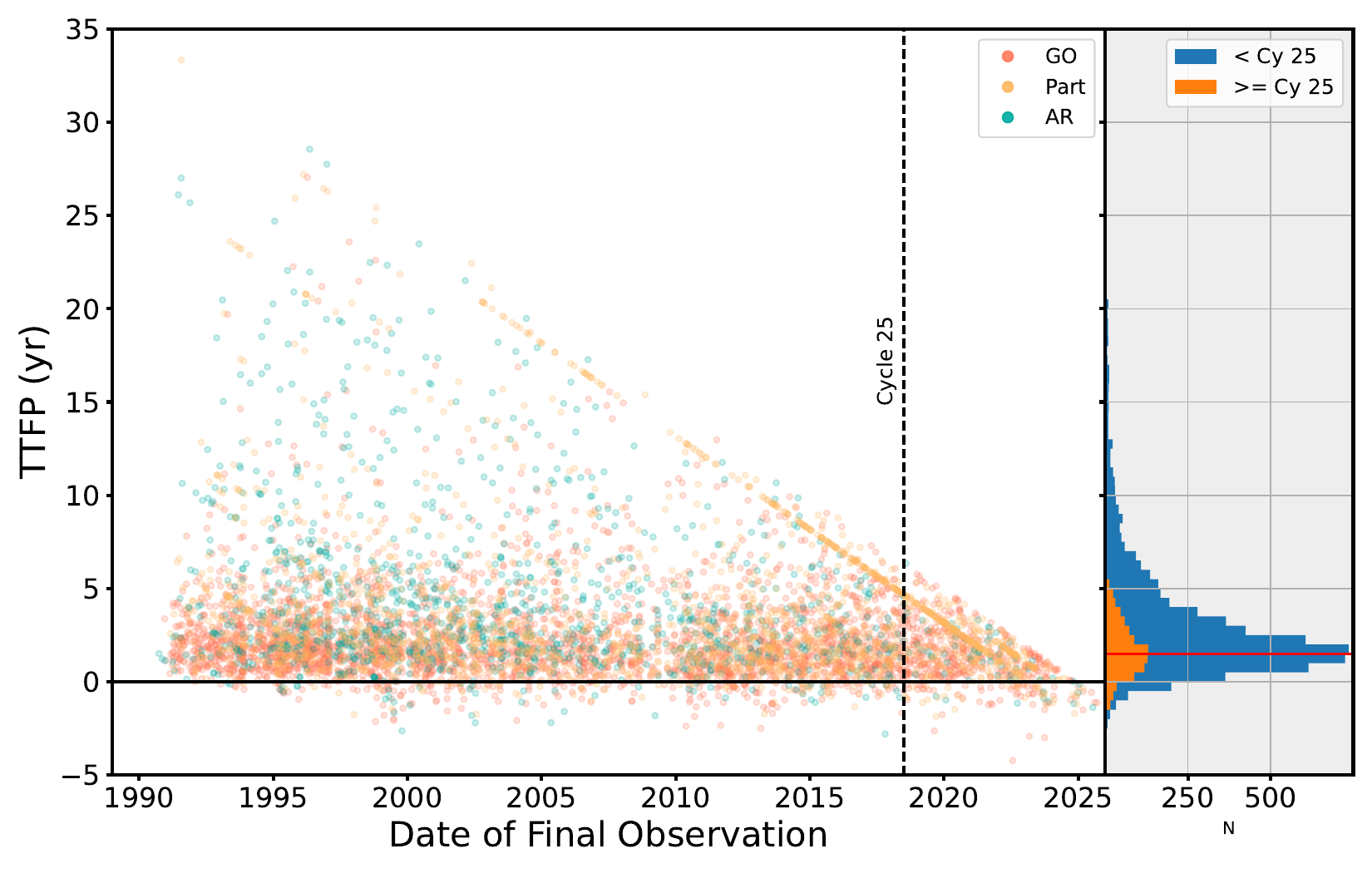}
\caption{Left panel: the time elapsed from the final observation in an \HST\ Observing program to the date of the first refereed publication (TTFP) vs. date of the final observation, color-coded by author category (as in Fig~\ref{fig:obsPubs}; see legend). The diagonal feature with a slope of $-1$ corresponds to an archival paper that used data from 1652 programs dating back to 2002 \citep{Kruk23}. The start of observing Cycle 25 (dashed line), when the maximum exclusive access period was revised from 12 to 6 months, is indicated (\textit{dashed line}). Right panel: histograms of the time lag for programs awarded prior to Cycle 25 (\textit{blue}) and afterward (\textit{orange}). A typical TTFP prior to Cycle 25 is about 1.5 yr (\textit{red line}) judging by the peak in the distribution. 
\label{fig:ttfp}}
\end{figure*}

Figure~\ref{fig:ttfp} shows the elapsed time between the conclusion of an observing program and the publication date of the first refereed paper (i.e., the time to first publication: TTFP), vs. the date of the final observation for the program. For this purpose we considered only those programs awarded through competitive peer-review (GO, SNAP, Treasury), as well as GTO and DD program categories (see Appendix~\ref{sec:progCategories} for category definitions). While data obtained from the categories we excluded (e.g., those related to calibration, engineering, on-orbit science verification, public outreach) often result in science publications, the incentives for prompt publication do not apply in the same way. The histogram in the right panel of Fig.~\ref{fig:ttfp} shows that, roughly speaking, a typical publication lag was about 1.5 years (with a skew toward larger lags).
In Cycle 25 the maximum EAP was shortened from 12 to 6 months. It is \added{clear} that GO papers are typically the first to be published, but if the first publication occured more than 5~yr after the last data become available then Archive or Part Archive papers tend to dominate. 
After the EAP was shortened the publication lag 
\added{appears to be shorter} 
by a few months. Negative lags simply mean that some authors published initial scientific results prior to the final observations for a program; science programs that include many targets, for example, lend themselves to incremental publications. The time to first publication includes the peer-review process and the journal publication process, which together often take a few months. Note however that authors often make their papers available on \href{https://arxiv.org/}{arXiv.org} upon acceptance, or several weeks earlier than the journal publication date, which would not be reflected in Fig.~\ref{fig:ttfp}.

\begin{figure}[h]
\plotone{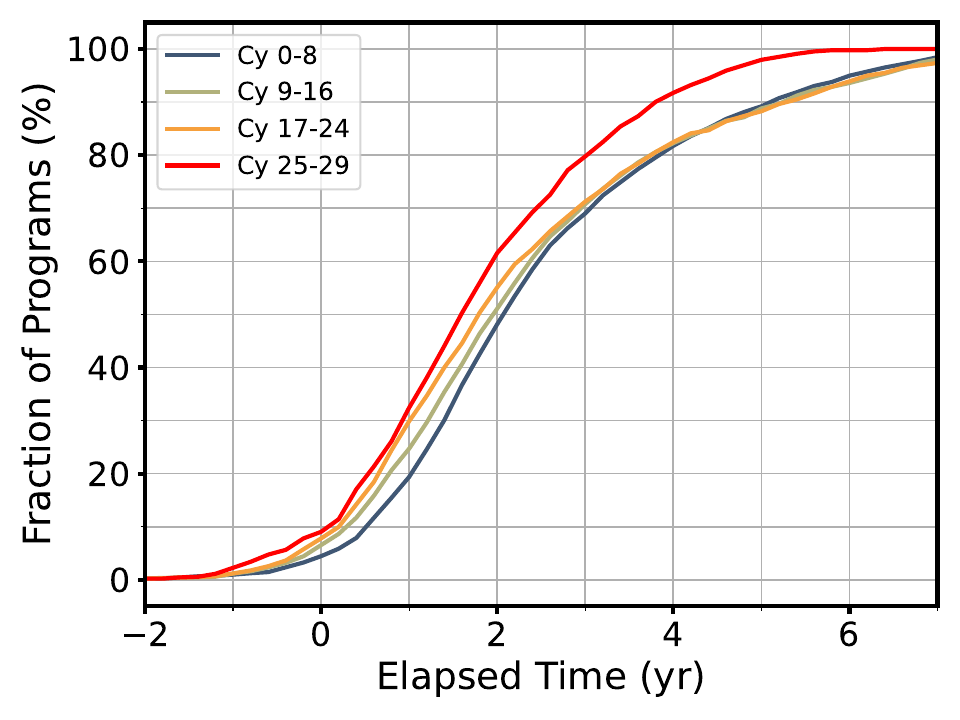}
\caption{\added{Cumulative fraction of first publications for ranges of \HST\ observing cycles (see legend) vs. the time elapsed from the final program observation to the publication date.} 
\label{fig:hstTtfpCurves}
}
\end{figure}

A closer look at the distribution of TTFPs over time (Figure~\ref{fig:hstTtfpCurves}) 
shows more clearly the effect of decreasing the maximum EAP for \HST. For this plot we excluded ``first publications" from Cycles 30 and later, and where publication dates that occurred 8 or more years after the data became available, to mitigate the cut-off effect. 
\added{
We divided the observing cycles into four sequential time periods to search for a temporal trend in the TTFP. The time elapsed to the first publication for 50\% of programs looks to have decreased slowly and systematically from about 2.1 yr for Cycles 0 through 8, to about 1.6 yr for Cycles 25 through 29. A first publication has appeared within about 3.8 yr for about 80\% of programs awarded in Cycles prior to 25; the shorter TTFP for Cycles 25 through 29 is unreliable because of truncation---i.e, these programs are too recent to measure long time lags. It is not obvious that the change in Cycle 25 to the maximum EAP had any discernable effect to what seems to be a slow decline in TTFP. This decline may instead be attributable to other factors, such as improved efficiency in the paper publication process, researchers familiarity with \HST\ data, or other factors.
}

Through \HST\ Cycle 29, where most data had been archived by mid-2022, about 6800 of the 7700 GO, GTO, SNAP, and DD observing programs (88\%) have resulted in at least one science publication. This is roughly consistent with the publication fraction curves in Figure 3-8, and indeed Fig.~\ref{fig:ttfp} suggests that the first publication for some programs appears many years after the observations become available. 

Occasionally discussions at various administrative levels (e.g., within NASA or within mission user committees) attempt to address the balance between the needs of the proposers who are awarded telescope time based upon competitive peer review (and who craft the observation plan) vs. archival users who would like to use new data for their science as soon as possible. For \HST\ at least, the bulk of the programs will result in many (and perhaps hundreds or thousands of) publications. Empirically the change in the duration of the EAP 
\added{appears not to have affected} the TTFP for \HST, 
though the TTFP is currently longer than the EAP by a factor of about 3. 

\subsubsection{Publication Authors}

Returning our focus to all MAST-hosted missions, we examined the authorship of science publications to shed some light on the breadth of the community of researchers. We fetched from ADS the author lists for each science publication of MAST-hosted missions. We attempted to homogenize the author names to aggregate publications of each first author, though this is a notoriously difficult challenge because of the variety of ways that authors choose to represent their names in publications. (ORCIDs do not solve this challenge for the oldest publications.) To this end we matched author surnames and initials, which for example properly associates the common case of ``Shaw, Richard" and ``Shaw, R." if they are the same person, but incorrectly associates ``Shaw, Richard" and ``Shaw, Robert" and treats ``Shaw, R. A." as a different person. \cite{Ness25} used a similar technique for analyzing authors in the \XMM\ bibliography. They point out that such mis-associations are likely to have only a modest affect of overestimating the number of papers per author while underestimating the number of different authors. 

\begin{figure}[h!]
\plotone{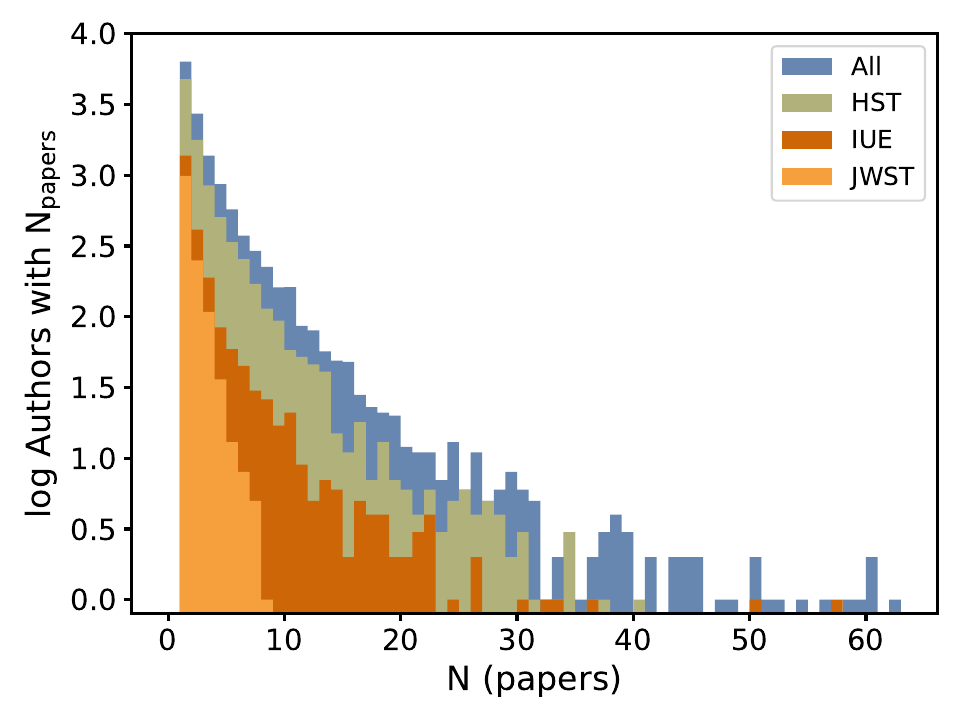}
\caption{Number of investigators who have published N papers as first author, over all MAST missions as well as selected missions (see legend). There were nearly 13,900 first authors over all missions. 
\label{fig:mastFirstAuthors}}
\end{figure}

There were over 45,000 unique authors of refereed science articles over all MAST-hosted missions published prior to 2025 Jan. Nearly 13,700 were first authors of science publications that published data from any MAST mission, with 9200 for \HST, just over 2300 for \IUE, and about 1400 for \JWST. Figure~\ref{fig:mastFirstAuthors} shows the number of investigators who have published a given number of papers as first author, segregated by select hosted mission. The histograms of first-author publications for all missions peak at $N=1$, which is consistent with a continuous flow of new first-authors throughout each mission lifetime. Each curve shows a roughly power-law decline, with a slope that flattens with the duration of the mission. Naturally, the observed maximum number of publications by a single first author also increases with mission lifetime. 

\section{Discussion\label{sec:discussion}}

Researchers in the international community have made prolific and imaginative uses of MAST data and services, as evidenced in the extensive bibliographies of refereed publications for the missions it hosts. Common interfaces to the MAST data holdings, robust back-end services, unifying databases of content, and high system reliability are among the features that make the impact of MAST more than just the sum of impacts from the individual missions. But there are many other factors, some of which are external, that contribute to the success that MAST enjoys today, and which directly affect the productivity and impact of the hosted mission data. Most of these same factors also contribute to the success of other astronomical archives. 

\subsection{Factors Affecting the Success of MAST}
\subsubsection{Funding}

While not the subject of this paper, NASA and its international agency partners have provided a strong, decades-long commitment for mission archives. This flows in part from a commitment to public access to taxpayer-supported data, and to a recognition of the potential for re-use of science data to leverage new science. This support comes in many forms, direct and indirect, including the construction of data processing pipelines, excellent calibrations, computing infrastructure, data analysis software, supporting databases, as well as direct support to investigators to carry out research. Support has also been available to continuously upgrade (over many generations) software, hardware, networking, and other infrastructure. We show in this paper many ways in which this investment over time in MAST has paid off handsomely. While missions launch, operate, and conclude, it is clear that a modest investment in their archives will result in a continuing return of science over the decades to come. 

\subsubsection{High Quality Data}

As mentioned in Sect.~\ref{sec:Intro}, the primary products in MAST are science-ready. Most have been produced by high quality science processing pipelines, with highly accurate calibrations. The provenance of the data (and the processing software) is very well documented, and for the Flagship missions the software is open-source and straightforward to customize for special science cases. The metadata, which describe (among other things) the footprint of the data in space, time, and energy, is also accurate, complete, and comprehensive. This facilitates search and discovery by researchers beyond those who originally obtained the data. It helps that mission data have been collected over a long time baseline (c.f. Fig.~\ref{fig:mastPubs2025}). Mission data from \IUE\ are still used today to complement data obtained from more recent missions such as \FUSE, \GALEX and \HST. The long mission life of \HST\ also ensures continuity and consistency of data, calibrations, and to inspire large, multi-observatory community surveys which have enduring archival value. Of course, the long life of the \HST\ observatory owes much to hardware upgrades and new science instruments that were installed through the first two decades of its lifetime. 

\subsubsection{Advances in Computing Infrastructure}

From the earliest days of the \HST\ archive (now MAST), which began science operations more than 35 years ago, compute power, storage capacity, network transfer speeds, etc. have improved by orders of magnitude. Supporting software (languages, web services, build tools, etc.) and database technology have all also grown at an amazing rate. 
\added{For the Flagship missions, cloud-based computing allows upgrades to processing software and calibrations to be applied within weeks rather than months, leading to improved science products. 
}
All of these technical and engineering advances were driven by commercial or government entities, and all are enabling technologies for MAST. Providing advanced programmatic and web user interfaces, advanced visualization and search tools, performant databases, and fast data retrieval make extensive archival research achievable and efficient. 

\subsubsection{Cross-Archive Collaborations}

\added{MAST staff participate in the development of community standards, tools, services, and software infrastructure, such as those developed within the International Virtual Observatory Alliance (IVOA), via the US component: the US Virtual Observatory Alliance. MAST also employs tools and web-based resources developed at partner archives and data centers, including our partner archives that mirror \HST\ and \JWST\ data: 
the Canadian Astronomy Data Center (CADC) and 
the European Space Astronomy Center (ESAC). 
MAST also collaborates with and uses tools and resources from:  
the Astrophysics Data System (ADS), 
the Centre de Donn{\'e}es Astronomiques de Strasbourg (CDS). 
the \textit{Chandra} Archive, 
the High Energy Astrophysics Science Archive Research Center (HEASARC), 
the InfraRed Science Archive (IRSA), 
the NASA Exoplanet Archive (NEA), and
the NASA/IPAC Extragalactic Database (NED). 
Collaboratively developed common standards, infrastructure and tools are critical to pan-chromatic research, within the MAST ecosystem and beyond. 
}

\subsubsection{Size Matters}

The total volume of data hosted at MAST for all missions currently exceeds 4 PB and close to a billion artifacts, not counting back-up safe-store resources. This volume is expected to grow very rapidly once the \textit{Nancy Grace Roman} mission (Akeson, et al. 2019)  is launched. MAST today is likely not the largest digital archive of astronomical data, nor is the size remarkable compared to that for certain other disciplines. MAST does, however, host archives for more than two dozen missions, with data mostly in the UV-Optical-IR wavelength regime, and many of which have enjoyed very high demand among the research community. While it is technically possible to search for and discover data in many digital astronomical archives worldwide, having data in one place that can be accessed through a common set of interfaces does facilitate research. And as the size of the community grows (as evidenced by the increasing publication rate for refereed journals), the demand for high quality archival data will also grow.

But indirect effects also matter. The back-end IT infrastructure that supports network connections, security, computation, web services, storage, user interfaces, etc. tends to be scaled to the most demanding mission in an archive. Access to other smaller, perhaps legacy mission data in the same archive benefits from this continued investment in infrastructure. One side effect of that investment is to make it easier to leverage opportunities for cross-mission research, such as for \HST\ with \IUE\ or \JWST; \TESS\ with \Kepler, and K2; and \GALEX\ and Pan-STARRS with all of them. 

\subsubsection{Telescope Time Allocation}

Apai, et al. (2010) argued that, for \HST, the strategy for allocating observing time has a direct bearing on the kinds of science programs that can be addressed, and the topic-based communities that can be served with a GO-oriented observatory. The \HST\ and \JWST\ missions take steps to ensure that the size of the observing time request is less of a determinant to success than the scientific merits. Thus, fainter targets, more ambitious science goals, or extreme urgency need not be a barrier to success. For the MAST, the range and diversity of science goals that were addressed in mission data only adds to the appeal for data re-use, and this is reflected in the metrics for productivity and impact for the various time allocation categories (see Sect.~\ref{subsec:pubEvol}). 

\subsection{Publication Metrics as Measures of Impact}

Several authors have employed the rate of published papers as a measure of productivity, and citations to those papers as a measure of impact for NASA missions or other observatories. Some (Crabtree 2008; Rots, et al. 2021; Scire, et al. 2022) have noted that these statistics based solely on numbers of publications and citations are incomplete or can be misleading, and depend, among other things, upon how carefully and consistently the bibliographies were constructed. We examine the power and limitations of the bibliometrics for MAST in the following sub-sections, but first we note some of the environmental factors that affect the publication metrics. 

\subsubsection{Size of the Research Community}

The size of the community that publishes research relates directly to the rate of refereed publications in the field. This is true for astronomy vs. physics (ADS shows that the latter has a publication rate $>10$ times that in astronomy) or within astronomy itself. The number of astronomy publications in the main journals (see Sect.\ref{subsec:pubEvol}) grew by a factor of 5.6 in the five decades to 2024, as have individual sub-fields over time (e.g., galaxy formation, cosmology, exoplanet studies). Since different missions may be more or less relevant to a given sub-field, the publication rates and therefore the citation rates will reflect that. For example, in MAST we see many more publications related to exoplanets for \Kepler, K2, and \TESS\ than we do for \HST. Similarly, research on solar-system bodies is more commonly based upon data from other observatories and planetary fly-by missions, rather than on \HST. This complicates comparisons of mission productivity, even if the missions are of the same class. 

\subsubsection{Discovery Space of the Mission}

The impact of a mission is likely higher where one or more aspects of the data are rare or in high demand—that is, where the discovery space is large. Missions with long in-service lifetimes, that cover large areas of sky, that obtain data in a new wavelength domain, that offer an large advantage in sensitivity or resolution, or that collect long-duration time-domain data on a large set of targets: all have high potential for generating large numbers of high-impact papers. 

\subsubsection{Impact Beyond Publications}

Deriving deeper insight into mission publication statistics is easiest when bibliographers have access to mission program data: which types of science programs were selected, how much observing time was awarded, to what extent program data were re-used, etc. This is not really possible for missions where the primary objective was a survey, nor where program information is not available, nor for missions that had a very brief in-service lifetime; these limitations apply to most missions hosted by MAST. Also, measuring the impact of surveys is itself challenging: if a team constructs a catalog from mission data (particularly if cross-matched with external catalogs) then publishes a catalog that subsequent investigators use, do papers that use the derived catalog count as mission papers? Currently for MAST they do not (unless the authors contribute the catalog to MAST for distribution), but this approach may have to be revisited as future missions like \textit{Roman} will produce voluminous catalog products, the impact of which will need to be measured. 

While paper and citation counts provide insight into mission impact, they are not perfect proxies. Mission archives can impact a research sub-field in ways that are difficult to measure. Archives are very good for understanding the state of observations of a class of astrophysical sources, but there is no real metric for understanding what inspires surveys, such as the Hubble Deep Field \citep{Williams96} or CANDELS \citep{Koekemoer11} that may transform an entire sub-field of research. Even a single image, or a few catalog entries may be enough to inspire a whole new way of thinking within a small sub-field of astronomy. Mission archives and the publications based upon them can also inspire professional conferences, the creation of teaching curricula, or of supporting proposals for new research. Publication of images or other artifacts in public media has the power to inspire future generations researchers or engender support among the tax-paying public. All such indirect impacts are of great benefit to the discipline of astronomy, but none are well measured by publication statistics. 

\subsection{Evolution of Publishing\label{subsec:pubEvol}}

\begin{figure}[h]
\plotone{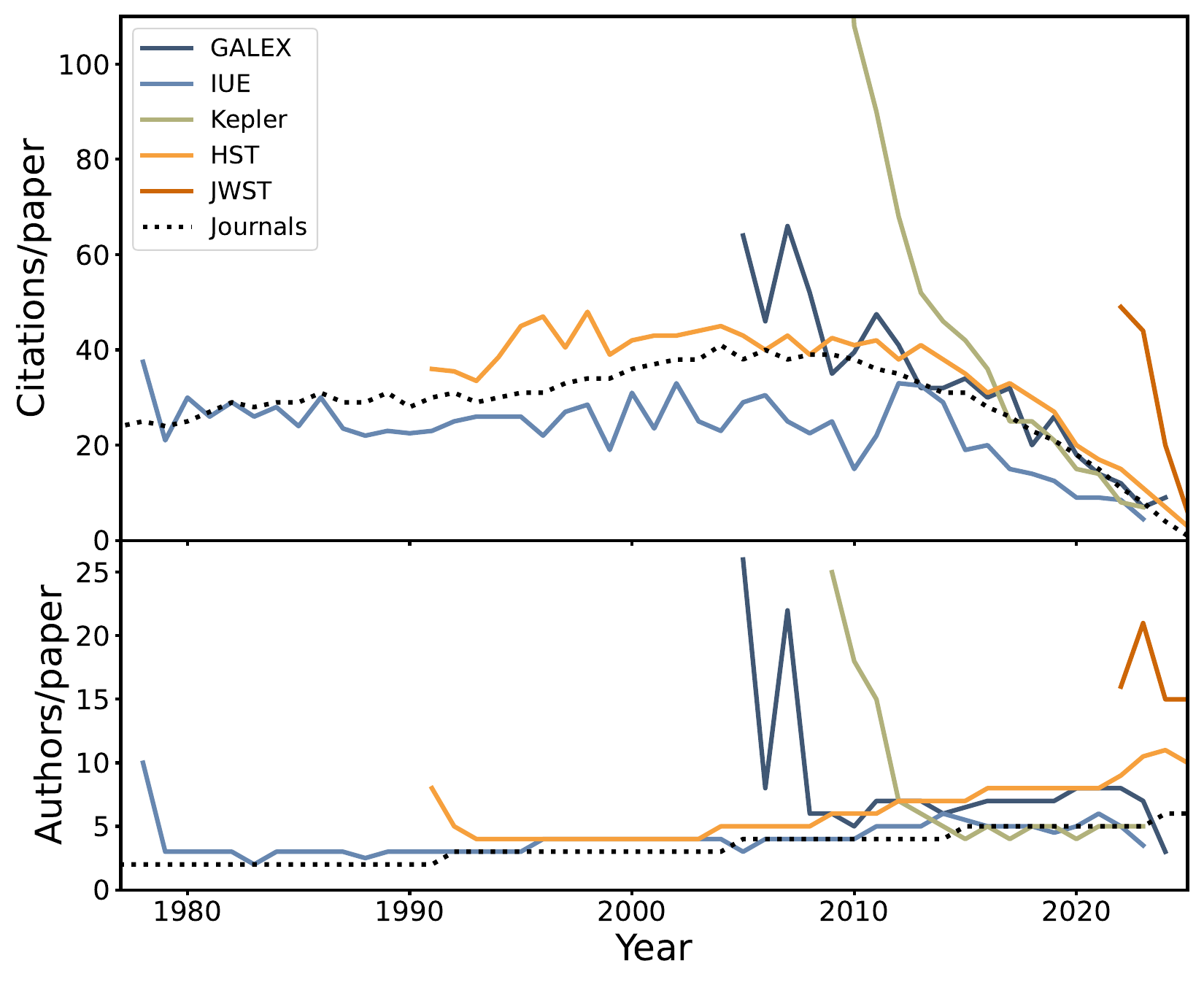}
\caption{Evolution of the median citations per paper (\textit{upper}) and median authors per paper (\textit{lower}) as a function of time for selected MAST missions (see legend), and for papers published in the main journals noted in Fig.~\ref{fig:papersByJournal}. 
\label{fig:authCites}}
\end{figure}

Authors of prior bibliometric studies in astronomy have remarked upon the increase over time of the number of authors per paper \citep{Abt81, Crabtree08}, the citation rate per paper \citep{Meylan04, Crabtree08} and a purported correlation between the number of citations and the number of authors \citep{Crabtree08}. We find no correlation whatever between numbers of co-authors and citations to MAST mission science papers. Publications with large numbers of co-authors appears to be a recent phenomenon: searches with ADS show that the first paper in astrophysics with more than 1000 co-authors appeared in 2017, and the first with more than 200 co-authors appeared within the last 15 years. Such papers are not typical, however. The publication statistics for MAST-hosted missions paint a different picture, as shown in Figure~\ref{fig:authCites}. While the median number of authors per paper in the primary journals (see Sect.~\ref{sec:corpora}) has increased steadily from one in 1945 to six in 2025, the cause does not appear to be related to the demands of publishing results from space missions, or at least not those hosted at MAST. What is more interesting is the noticeably larger number of authors per paper in the first year or two of a mission. In early mission papers, topics often include the on-orbit performance of the new observatory, which naturally would include large teams of people as co-authors. This is particularly true of survey missions such as \GALEX, \Kepler, and \TESS. Most missions more or less track the trend of journal articles generally after the first two years. It is too soon to draw firm conclusions about the co-author numbers and the citation rate for \JWST. We note that the instruments on \JWST\ have a large number of available configurations and data-taking modes, and observing time is in very high demand. We suspect that large teams are an advantage for making a compelling case for observing time, preparing the typically complex observing programs, and extracting all of the science from the observations. 

We find that the median citation rates per paper per year in the primary journals has also increased steadily in time, from one in 1945 to nearly 40 in 2010, but then declines afterward (upper panel of Fig.~\ref{fig:authCites}). We attribute this decline between 2010 and the present day to the time needed for papers to accumulate the bulk of their citations (see Sect.~\ref{subsec:citesMissionPubs}). This implies that conclusions in prior bibliometric studies about impact which are based upon aggregate citation rates may not have fully accounted for the long timescale for citations to accumulate. For MAST missions, the median citation rates generally track that for papers in the primary journals, again with two exceptions. Survey missions such as \Kepler, \GALEX, and \TESS, tend to have elevated citation rates at the very outset (perhaps because the mission definition team members are well prepared for the first data to appear); the rates for \JWST, while extremely preliminary, suggest an outsized pending impact on the research community.

\subsection{Mission Productivity}

It is perhaps no surprise that missions hosted by MAST show a high or very high publication rate (see Fig.~\ref{fig:mastPubs2025}) during their operational phase or shortly afterward, but it is remarkable that the publication rate for most missions remains significant for years or even decades beyond the end of operations. This is particularly true for missions with a large survey component (\FUSE, \GALEX, \Kepler, Pan-STARRS), but even for a non-survey missionn like \IUE, science publications have continued for three decades since it was decommissioned. While the post-operations papers are mostly archival research, it is remarkable that archival research accounts for such a large component of productivity for the active missions \HST\ and \JWST, and if \HST\ is any guide, archive research eventually dominates mission productivity. 

\subsubsection{Productivity by Program Category\label{subsubsect:prodByProg}}

We noted in Section~\ref{sec:impact} the remarkable number of published papers and citations to those papers for the \HST\ mission. We also noted that a modest fraction of all papers, for multiple missions, accounts for an outsized fraction of all citations. Part of the reason that some papers have outsized impact flows from policies that govern the types of observing programs and the time allocation categories offered by the \HST\ and \JWST\ missions. Appendix~\ref{sec:progCategories} summarizes the current proposal categories that are offered; we note that these categories and their definitions have evolved somewhat over time. Table~\ref{tab:pubStatsByTime} shows the productivity and impact by time allocation category where observations are mostly complete (through \HST\ Cycle 29 and \JWST\ Cycle 2). Note that Treasury programs usually have large or very large allocations of observing time, but have a broader scientific purpose. Following the category, successive table columns give the number of programs with that category, the median of the number of papers that used data from that program, and the median of accumulated citations to the papers associated with each program. We also computed for each program the ratios of the number of papers to the time allocation, and the citation count to the time allocation; the last two columns give the median values of these quantities for each category. It is important to note that the units of time allocation are orbits for \HST\ but hours for \JWST; the time allocations can include the time needed for instrument configuration, exposure overheads and small-angle maneuvers for dithered and tiled exposures. These factors makes comparisons of efficiency meaningful between program categories for a given mission, but less so between missions. Some \HST\ observing proposals were split into multiple programs for the sake of scheduling and observing efficiency. We aggregated such cases under the lowest-numbered program so that paper counts and citations (which would be highly correlated) would not be multiply counted in our statistics. Finally, we re-categorized a few \HST\ Director's Discretionary (DD) programs as Treasury, either because they fulfilled the attributes of the Treasury category even though that category did not exist when these programs were created, or because certain otherwise qualifying multi-cycle programs were difficult to propose other than through DD allocations. 

\begin{deluxetable}{lrrrrr}[h!] 
\tabletypesize{\footnotesize}
\tablewidth{0pt} 
\tablecaption{Publication Statistics per Observing Program Time Allocation Category\label{tab:pubStatsByTime}}
\tablehead{
  \colhead{}         & \colhead{Program} & \colhead{Median} & \colhead{Median} & \colhead{Papers} & \colhead{Citations} \\
  \colhead{Category} & \colhead{Count} & \colhead{Papers} & \colhead{Citations} & \colhead{per Alloc.} & \colhead{per Alloc.}
}
\startdata
\cutinhead{\HST}
Treasury\tablenotemark{a} &   49 & 80 & 3281 & 0.437 & 28.3 \\
Joint    &  353 &  4 &  118 & 0.667 & 19.3 \\
Large    &  308 & 15 &  751 & 0.133 &  7.1 \\
Medium   &  468 &  8 &  340 & 0.174 &  7.7 \\
Small    & 1210 &  6 &  256 & 0.276 & 11.3 \\
V. Small & 4146 &  3 &  109 & 0.667 & 21.5 \\
\cutinhead{\JWST}
Large    &  17 &  19 &  735 & 0.110 &  5.4 \\
Medium   &  50 &   8 &  244 & 0.134 &  3.4 \\
Small    & 122 &   3 &   42 & 0.087 &  1.4 \\
V. Small & 164 &   2 &   31 & 0.176 &  3.0 \\
\cutinhead{Cross-Mission}
HLSP     & 249 & 317\tablenotemark{b} &   54 & \nodata & \nodata \\
\enddata
\tablenotetext{a}{A few early and large DD programs, including the Hubble Deep Field and the Ultra-Deep field, were re-categorized as Treasury for this analysis.}
\tablenotetext{b}{Paper count for HLSP collections is a total.}
\end{deluxetable}

Treasury and Large programs account for an outsized fraction (per program), of both published science papers and citations to those papers. This is consistent with their purpose: these proposals are evaluated in part on their potential to generate data useful for many scientific purposes. These programs were often provided funding for investigators to produce high-level science products (often, combined images) that are released through MAST for use by the community. In terms of their contribution to \HST\ science productivity, impact, and data re-use for follow-on archival research, these program categories have succeeded magnificently. Indeed, one Treasury program \citep[CANDELS: ][]{Koekemoer11} accounts for roughly 8\% of all \HST\ citations. It would be a mistake, though, to conclude that productivity and impact can be maximized by granting observing time only to very large programs. When viewed in terms of the observing time allocations, the Very Small and Joint programs are more efficient at producing papers and nearly as efficient at producing citations. These conclusions roughly mirror the \cite{Apai10} findings, except that Joint programs had only just been implemented, and our updated metrics for the Treasury programs are similar to or exceed that of any category. In the end, though, the impact of any observatory is best measured by scientific results, and this is not the same as papers and citations to papers. It is worth acknowledging that the impact of a long mission can also be measured by support for a range in observing time allocations, with the recognition that different kinds of time allocations support different kinds of achievable science. 

We take special note of two types of categories listed in Table~\ref{tab:pubStatsByTime}. Joint programs are those where new data are obtained from two or more observatories to achieve the science goals. (See Appendix~\ref{sec:progCategories} for a list of cooperating observatories.) Proposals for such programs are evaluated by only one peer-review panel, which if approved means observing time is allocated on all requested observatories. We list the publication statistics for \HST\ Joint programs that were awarded through Cycle 29; the number of \JWST\ Joint programs awarded through Cycle 2 are too few in number for summary statistics to be meaningful. Joint programs are relatively efficient at generating papers and citations to those papers, per unit of observing time. In spite of the small number of such programs, they are also evidence for cross-mission synergy beyond the missions hosted by MAST. The other category, High-Level Science Product (HLSP) collections, are not a type of observing category per se: most of the data products are derived from other science products hosted by MAST. Yet the products (e.g., photometric catalogs) are also often derived from multiple missions, sometimes including products from missions or observatories not hosted by MAST. Thus, HLSP collections are also a measure of cross-mission synergy, and papers based upon these products are well cited.  

We tabulate the publication metrics for \HST\ and \JWST\ by time allocation category in Table~\ref{tab:pubStatsByProp}, where we show the fraction of programs, the median of the papers produced, the citations for each program type, and the fraction of observing time allocated. The relatively high median citation rate for \HST\ Survey program papers is striking, especially since the genesis for this program type was largely to increase the efficiency of the observatory during times when no primary programs could be scheduled. It is too early to assess the success of this program type for \JWST, given the small number of \JWST\ Survey programs awarded through Cycle 2 that have resulted in a publication. The most remarkable result for \JWST\ is the high productivity and impact of the Early (ERO and ERS) programs. The data were released early to the public in part to foster archival research at an early stage. This appears to have been a stunning success. 

\begin{deluxetable}{lrrrr}[h!] 
\tabletypesize{\footnotesize}
\tablewidth{0pt} 
\tablecaption{Publication Statistics per Observing Program Time Allocation Category\label{tab:pubStatsByProp}}
\tablehead{
  \colhead{}     & \colhead{Program} & \colhead{Median} & \colhead{Median}    & \colhead{Time Alloc,} \\
  \colhead{Type} & \colhead{(\%)}    & \colhead{Papers} & \colhead{Citations} & \colhead{(\%)} 
}
\startdata
\cutinhead{\HST}
GO     & 77.5 & 4 & 145 & 67.8 \\
GTO    & 10.0 & 5 & 145 &  3.7 \\
DD     &  7.9 & 3 & 145 &  5.8 \\
Survey\tablenotemark{a} &  5.4 & 7 & 327 & 22.7 \\
\cutinhead{\JWST}
GO     & 55.0 &  2 &  37 & 53.4 \\
GTO    & 28.5 &  3 &  44 & 26.3 \\
COM    &  8.2 &  2 &  36 &  6.9 \\
Early  &  5.7 & 18 & 586 &  5.7 \\
DD     &  5.1 &  4 & 136 &  1.6 \\
Survey\tablenotemark{a} &  0.6 &  1 &  10 &  3.0 \\
\enddata
\tablenotetext{a}{Survey program targets are approved but observations are not guaranteed. The typical completion fraction is $<50$\%.}
\end{deluxetable}

\cite{Rots12} argue persuasively that metrics beyond counts of papers and citations are more meaningful, and suggest metrics related to the fraction of data that are published, the fraction of observing time that results in a publication, the speed of publication after the data are made available, and the number of citations that are significant to science in the citing paper. We derived most of these quantities in this work and agree with the aims, but there are practical challenges. For instance, the fraction of available observing time is not clear-cut: \HST\ and \JWST\ observations for different programs can be observed in parallel. Also, different types of constraints and overheads affect observatories differently. 
For instance, the \Chandra\ orbit allows for long integrations relative to set-up time (which is very helpful for a photon-starved observatory), but for \HST\ there is a good deal of unusable time because of Earth occultation during the \HST\ orbit. 
Similarly the ``warm" portion of the \Spitzer\ mission greatly affected the integration times because of the elevated IR background. For survey missions, the fraction of productive observing time is hardly relevant. 
With regard to citations, it is impractical to discern when citations are scientifically significant (rather than incidental), and we worry that this would be a distinction without a difference unless one is prepared to believe that citations to \HST\ are more (or less) on point relative to some other mission. 
The results presented here do take observing efficiency into account, and stand on their own in the context of a single mission, and to a limited extent in aggregate over all MAST missions. But detailed comparisons of publication metrics between individual observatories is fraught and inexact at best. 

\section{Conclusions\label{sec:conclusions}}
\subsection{The High Impact of MAST Missions}

The mission data hosted in the MAST archive supports a large astronomical research community, and has enabled the production of over 37,000 publications on a wide range of astronomical topics. We have constructed bibliographies of refereed articles for each mission from nearly 40 professional journals, and have analyzed publication statistics for articles that made science usages of data. We found that our discovery of articles classified as Science is at least 90\% complete (95\% for the Flagship missions), with very good classification accuracy. The publication rate of Science articles over the last 50 years shows that most MAST missions have had very high productivity during their in-service lifetimes, and remain very productive for years or decades afterward. The publication rate for \JWST\ Science articles is extraordinary, and shows the steepest post-launch growth of any mission hosted by MAST. 
The median rate of citations to Science papers is in keeping with that for general articles in the professional journals, but for articles based on Treasury (large) observing programs it is remarkable, suggesting in aggregate a high impact on astronomy as a discipline. Most of the impact, as measured by citations, comes from a small fraction of articles within each mission, some of which have extraordinary citation rates. 
Individual articles accumulate citations over a timescale of a decade or more, which is also in keeping with refereed articles in the journals. We found that the rate at which citations grow varies widely among articles. Indeed for some rare, highly cited articles the citation rate is still accelerating. 

We found that the first publication for a small fraction \HST\ observing programs appears a decade or more after all program data have been collected; the fact that not all of these were classified as archival suggests that our classification criteria likely leads to over-counting GO papers. For many programs, archival papers were published soon after the data are made public. We find that most authors publish only a single Science article as first author, and the number of researchers with multiple first-author articles falls off roughly as a power law. About 10\% of articles analyze data from multiple MAST missions, but cross-mission synergy is difficult to measure in the absence of comparable bibliographic information from missions and observatories not hosted by MAST. 

We examined the bibliographies of the active missions \HST\ and \JWST\ in greater detail by cross-referencing our bibliography with the respective observing program databases. We found that the rate of published archival papers for \HST\ exceeded those published by the original observing team (GO) within a decade of the start of science operations; early indications for \JWST\ tempt us to speculate that archival papers could dominate much sooner. Currently the rate of \HST\ archival papers exceeds those by GOs by a factor of more than 3. The production of papers for any given \HST\ program can extend for decades, but a few very large (Treasury) programs that were intended to have very broad scientific appeal have generated hundreds or thousands of papers each, with tens- to hundreds of thousands of citations per program. For \HST\ programs in general, a first publication appears within 1.5 yr for 50\% of programs, and within 3.8 yr for 80\% of programs. 
We also found that \added{the time to first publication for \HST\ observing programs has slowly but steadily declined since the start of science operations, which may be indicative of greater efficiencies in data analysis and the publication process.}

The Flagship mission program information allowed us to characterize the productivity and impact of programs by allocation of observing time, and by the program type. 
We found that very large Treasury programs, programs with very small allocations of time, and Joint observatory programs ranked highest in papers and citations per unit of observing time. But the Treasury programs seem uniquely capable of generating the greatest level of follow-on science through data re-use. The productivity and impact for most program types was comparable, but the \HST\ Survey programs and the \JWST\ Early Release Science and First Images programs stood out for generating proportionately large numbers of papers and citations. 

\subsection{The Next Chapter of the MAST Bibliography}

\added{
We expect that MAST staff will continue adding papers to the existing mission bibliographies, and the growth in papers during the past 15 years compels us to do so much more efficiently. MAST will also track publications from new hosted missions, especially \textit{Roman Space Telescope}\ which is scheduled to launch in 2026. MAST is also beginning to host data from small, community-contributed missions.\footnote{\url{https://archive.stsci.edu/new-mission-partnerships-with-mast}}
} 

\subsubsection{Maintaining the Bibliography}

As we have shown here, mission bibliographies contain a wealth of information on the productivity and scientific impact of an observatory on the discipline of astronomy, as well as insights about the associated community of researchers. They also help to quantify the effects of observatory initiatives and policy choices. A mission bibliography requires a modest but continuing investment in staff time to review the literature, to create, maintain, and update supporting software and data systems, and to analyze the results. Yet they are not perfect: science articles can be missed or usage mischaracterized, some authors fail to provide proper attribution of facilities or data sources, and paper and citation counts are not perfect proxies for science impact. Some shortcomings can be minimized, but not without significantly greater cost in staff time. As a result, some established observatories may find the cost to create and maintain a bibliography prohibitive. 

We have begun to develop a new bibliographic review process that leverages artificial intelligence \citep[e.g.,][]{AmadoOlivo25,Wu25}, with the aim of automating as much of the classification work as practical. Initial results suggest it should be very good at identifying mission call-outs, and shows promise for classifying data usage accurately. There are many potential advantages of such an approach, beyond lowering cost. If successful, we would have the opportunity to reduce substantially any missed mission call-outs in previously reviewed MAST publications, and use machine classifications to identify cases where historical classifications of data use may be suspect. If successful, AI-assisted bibliographic review will boost our overall accuracy and reliability to an even higher level. Finally, one of the current barriers to comparing measures of impact for missions beyond those hosted at MAST is the imperfect alignment of criteria for science usage of mission data (Lagerstrom, et al. 2012). With machine classifications it should be possible to classify data usage from many more NASA missions or ground-based observatories using common and consistent criteria. Thinking further ahead, a perhaps more appropriate use of AI would be to ask questions relating to the scientific impact directly of the literature. We believe there is a whole frontier in this area to explore with AI. 

\subsubsection{Measures of Impact}

As with prior works, we described various limitations to using aggregate publication and citation rates as proxies for scientific productivity and impact for a mission. Coupled with details of individual observing programs it is possible to use mission bibliographies to characterize more deeply the impact of missions on astronomy. 
Yet there are still limitations to the kinds of analysis that might be useful, such the number and nature of catalog searches or cross-matches by investigators that will become much more commonplace in the next few years \citep[][track explicit use of \XMM\ primary mission catalogs, for example]{Ness25}. 
For catalogs, which themselves are becoming much more complex, impact would likely have to be measured in the context of multiple missions or observatories, for instance between the catalogs of \textit{Gaia}, the Rubin observatory, and the \textit{Roman} mission. The challenges are in part technical (e.g., how does one track cross-matches or sample selection among separate, cloud-based catalogs?), semantic (what does it mean to measure impact if the output is only indirectly referenced in a published paper?), and cultural (what are the expectations and guidelines from the publishers for authors to explain the details of their data usage?). These and other challenges await the next generation of bibliographers. 


\begin{acknowledgments}
We gratefully acknowledge the extensive and ongoing support from staff of the SAO/NASA Astrophysics Data System (ADS), funded by NASA under Cooperative Agreement 80NSSC21M00561. We are also thankful for the earlier work of J. Lagerstrom and E. Fraser, who formed the original basis of the bibliography work and establishing the standard we use for the science paper classification. The authors also acknowledge the contributions of team members C. Mesh, O. Oberdorf, S.-A. Tseng, and G. Wallace, who contributed to the software development and maintenance before departing STScI; G. Wallace passed away before this work was published. 
\added{This work made use of the following software packages: 
\texttt{matplotlib} \citep{Hunter:2007}, 
\texttt{numpy} \citep{numpy}, 
\texttt{pandas} \citep{mckinney-proc-scipy-2010}, and 
\texttt{python} \citep{python}.
Software citation information aggregated using \texttt{\href{https://www.tomwagg.com/software-citation-station/}{The Software Citation Station}} \citep{software-citation-station-paper,software-citation-station-zenodo}.
}
This research was supported by NASA contract NAS5-26555 to the Space Telescope Science Institute; funding for the bibliography work from 2005--2010 was provided by NASA grant NNG05-GF75G. 
\end{acknowledgments}




\facilities{MAST, ADS}


\clearpage
\appendix
\section{MAST-hosted Missions}\label{sec:mastMissions}

Table~\ref{tab:mastMissions} lists the Missions where science-ready data products are hosted at MAST, along with some alternative names for the missions that have appeared in the literature (searches for mission identifiers are case-insensitive). The table also gives the totals of refereed science publications through 2023 April (through 2024 December for \HST\ and \JWST), and the number of citations and the median number of citations to those papers as of late 2025.

\begin{deluxetable}{lllrrr}[h!] 
\tabletypesize{\footnotesize}
\tablewidth{0pt} 
\tablecaption{Missions Hosted at MAST\label{tab:mastMissions}}
\tablehead{
    \colhead{} & \colhead{} & \colhead{} &
    \colhead{} & \colhead{Total} & \colhead{Median} \\
    \colhead{Mission} & \colhead{Formal Name} & \colhead{Alternate Names} &
    \colhead{Publications} & \colhead{Citations} & \colhead{Citations} 
}
\startdata
BEFS\tablenotemark{a} & Berkeley Extreme and Far-UV Spectrometer & Berkeley Spectrometer & 68 & 1958 & 18 \\
\Copernicus & \Copernicus & OAO-3 & 193 & 21,536 & 25 \\
\EUVE       & \textit{Extreme Ultraviolet Explorer} &  & 499 & 21,625 & 26 \\
\FUSE       & \textit{Far Ultraviolet Spectroscopic Explorer} &  & 966 & 47,911 & 28 \\
\GALEX\tablenotemark{b} & \textit{Galaxy Evolution Explorer} &  & 3214 & 203,208 & 29 \\
HPOL\tablenotemark{c} & Half-wave Spectral Polarimeter &  & 44 & 1675 & 33 \\
 & & & & & \\
\HST\tablenotemark{d,e} & \textit{Hubble Space Telescope} & \textit{Hubble} & 22,604 & 1,442,980 & 33 \\
HUT\tablenotemark{c} & Hopkins Ultraviolet Telescope &  & 194 & 10,087 & 26 \\
IMAPS\tablenotemark{a} & Interstellar Medium Absorption & & 33 & 1104 & 20 \\
 & Profile Spectrograph & &  &  & \\
\IUE       & \textit{International Ultraviolet Explorer} &  & 5575 & 286,497 & 25 \\
\JWST\tablenotemark{d} & \textit{James Webb Space Telescope} & \textit{Webb} & 1363 & 68,411 & 27 \\
 & & & & & \\
\Kepler\tablenotemark{b} & \textit{Kepler Space Telescope} &  & 2211 & 159,677 & 34 \\
K2\tablenotemark{b}  & \textit{Kepler} extended mission &  & 700 & 29,255 & 25 \\
Pan-STARRS & Panoramic Survey Telescope and & Pan-STARRS1, PS-1, and & 1463 & 73,618 & 22 \\
           & Rapid Response System & variants without the hyphen &  &  &  \\
\TESS\tablenotemark{b} & \textit{Transiting Exoplanet Survey Satellite} &  & 1363 & 38,340 & 17 \\
TUES\tablenotemark{a} & Tubingen Ultraviolet Echelle Spectrometer & & 26 & 673 & 18 \\
 & & & & & \\
UIT\tablenotemark{c} & Ultra-violet Imaging Telescope &  & 112 & 3927 & 53 \\
WUPPE\tablenotemark{c} & Wisconsin Ultraviolet Photo-Polarimeter &  & 44 & 1657 & 31 \\
         & Experiment &  &   &   &  \\
HLSP\tablenotemark{f} & High Level Science Products &  & 315 & 38,313 & 55 \\
\enddata
\tablenotetext{a}{Flown aboard the Space Shuttle as a part of the Orbiting Retrievable Far and Extreme Ultraviolet Spectrometers (ORFEUS) payload.}
\tablenotetext{b}{Primarily a survey-oriented mission.}
\tablenotetext{c}{Flown aboard the Space Shuttle as a part of the ASTRO Observatory.}
\tablenotetext{d}{The acronyms and full names for each science instrument aboard the Flagship missions are also searched.}
\tablenotetext{e}{The names of special \HST\ advanced products: HLA and HSC, which are hosted by MAST, are also searched.}
\tablenotetext{f}{HLSP in this context is really a collection of community-contributed data collections.}
\end{deluxetable}

MAST serves certain data products from other missions and surveys which are not listed in Table~\ref{tab:mastMissions}, and where publications are not included in the this analysis.
\begin{itemize}
    \item The ESA \textit{XMM-Newton} Optical Monitor
    \item Data from various surveys within the Sloan Digital Sky Survey, which were being migrated to MAST starting in 2024
    \item Data from \textit{Gaia}, which is mirrored from ESA
    \item Certain data from the Extrasolar Planet Observations and Characterization (EPOCh) project
    \item Data from the UVOT instrument on the \textit{Neil Gehrels Swift Observatory}
    \item Data from the VLA Faint Images of the Radio Sky at Twenty-cm (VLA-FIRST) survey
\end{itemize}

\section{Flagship Observing Program Categories} \label{sec:progCategories}

Both the \HST\ and \JWST\ missions solicit(ed) proposals for observing time in multiple categories, which are described in Table~\ref{tab:obsCategories}. Different categories carry different expectations with respect to the allocated amount of observing time, the nature of the peer review process (usually through the Telescope Allocation Committee: TAC), and the maximum duration of the Exclusive Access Period (EAP). This table provides only a very brief description of each type; for details see the Call for Proposals for each mission. 

\begin{deluxetable}{llll}[h!] 
\tabletypesize{\footnotesize}
\tablewidth{0pt} 
\tablecaption{Observing Program Categories\label{tab:obsCategories}}
\tablehead{
    \colhead{} & \colhead{} & \colhead{Review} & \colhead{} \\
    \colhead{Type} & \colhead{Formal Name} & \colhead{Type} & \colhead{Description} 
}
\startdata
GO     & General Observer & TAC & Programs for general science are categorized by the \\
       &                  &     & size of the observing request \\
GTO    & Guaranteed Time Observer & NASA & Awards of telescope time in early cycles to teams who \\
       &                  &     & built one of the science instruments or developed other \\
       &                  &     & major observatory components \\
Survey & (Snapshot) TAC   & TAC & Proposers request separate, short-duration observations \\
       &                  &     & of a list of sources, which are observed only if no prime \\
       &                  &     & observations can be scheduled. There is no guarantee \\
       &                  &     & that any sources will be observed; completion fractions \\
       &                  &     & are generally less than 33\%. \\
Treasury & Treasury       & TAC & Larger programs that are designed to obtain \\
       &                  &     & observations with high potential for to solve multiple \\
       &                  &     & scientific problems. Observers are expected to produce \\
       &                  &     & and make available advanced data products with high \\
       &                  &     & potential for re-use by the community. \\
Joint  & Joint (Multi-observatory) & TAC & Programs that require new observations (of the same \\
       &                  &     & targets) from multiple reciprocal observatories to \\
       &                  &     & achieve the science goals. Proposals are peer reviewed \\
       &                  &     & by the observatory TAC associated with the prime \\
       &                  &     & science to avoid multiple jeopardy. Reciprocating \\
       &                  &     & observatories are currently: \Chandra, \HST, \JWST, \\
       &                  &     & NOIRLab, NRAO, \TESS, \textit{XMM-Newton} and, for \\
       &                  &     & \JWST only: \added{ALMA and} NASA-Keck. \\
DD     & Director's Discretionary Time & Special external & Programs of usually smaller size where the need for \\
       &                  & review & observations is urgent, or that significantly accelerate \\
       &                  &     & scientific discovery. \\
ERS    & Early Release Science & TAC & Part of DD time allocations, these programs were \\
       &                  &     & obtained in the first few months of \JWST\ Cycle 1 to  \\
       &                  &     & increase the number and breadth of public datasets. \\
ERO    & Early Release Observations & NASA & Also known as \JWST\ First Images, these programs \\
       &                  &     & were selected to demonstrate the science capabilities of \\
       &                  &     & \JWST\ before and just after the end of Commissioning. \\
\enddata
\end{deluxetable}


\clearpage

\bibliography{MAST_Biblio}{}
\bibliographystyle{aasjournalv7}



\end{document}